\documentclass[]{spie}  %>>> use for US letter paper
\usepackage{amsmath,amsfonts,amssymb}
\usepackage{graphicx}
\usepackage{lscape}
\usepackage{subfig}
\usepackage{xcolor}
\newcommand\crule[3][black]{\textcolor{#1}{\rule{#2}{#3}}}
\definecolor{blue}{RGB}{60, 145, 230}
\definecolor{orange}{RGB}{226, 132, 19}
\definecolor{green}{RGB}{13, 171, 118}
\definecolor{pink}{RGB}{206, 123, 145}

\usepackage[colorlinks=true, allcolors=blue]{hyperref}
\usepackage[version=4]{mhchem}
\usepackage{array, multirow}
\title{Atmospheric retrievals for LIFE and other future space missions: the importance of mitigating systematic effects}

\author[a]{Eleonora Alei}
\author[a]{Bj\"{o}rn S. Konrad}
\author[b]{Paul Molli\`ere}
\author[a]{Sascha P. Quanz}
\author[a]{Daniel Angerhausen}
\author[a]{Mohanakrishna Ranganathan}
\author[c]{the LIFE collaboration}
\affil[a]{ETH Zurich, Institute for Particle Physics \& Astrophysics, Wolfgang-Pauli-Str. 27, 8093 Zurich, Switzerland}
\affil[b]{Max-Planck-Institut f\"ur Astronomie, Königstuhl 17, 69117 Heidelberg, Germany}
\affil[c]{\url{http://www.life-space-mission.com}}
\authorinfo{Further author information: E.A.: E-mail: elalei@phys.ethz.ch}

\begin{document} 
\maketitle

\begin{abstract}
 Atmospheric retrieval studies are essential to determine the science requirements for future generation missions, such as the Large Interferometer for Exoplanets (LIFE). The use of heterogeneous absorption cross-sections might be the cause of systematic effects in retrievals, which could bias a correct characterization of the atmosphere.  In this contribution we quantified the impact of differences in line list provenance, broadening coefficients, and line wing cut-offs in the retrieval of an Earth twin exoplanet orbiting a Sun-like star at 10 pc from the observer, as it would be observed with LIFE. We ran four different retrievals on the same input spectrum, by varying the opacity tables that the Bayesian retrieval framework was allowed to use. We found that the systematics introduced by the opacity tables could bias the correct estimation of the atmospheric pressure at the surface level, as well as an accurate retrieval of the abundance of some species in the atmosphere (such as \ce{CO2} and \ce{N2O}). We argue that differences in the line wing cut-off might be the major source of errors. We highlight the need for more laboratory and modeling efforts, as well as inter-model comparisons of the main radiative transfer models and Bayesian retrieval frameworks. This is especially relevant in the context of LIFE and future generation missions, to identify issues and critical points for the community to jointly work together to prepare for the analysis of the upcoming observations.  
\end{abstract}


\section{INTRODUCTION}
\label{sec:intro}  % \label{} allows reference to this section

At the golden age of exoplanetology, the community is starting to focus on finding life in the universe. Terrestrial exoplanets, predicted to be very abundant, are the most likely candidates to be potentially habitable. However, these are currently very challenging to observe and characterize. Various new mission concepts are being designed to achieve this goal: from visible and near-infrared missions like the Habitable Exoplanet Observatory (HabEx)\cite{Gaudi2020} and the Large Ultraviolet Optical Infrared Surveyor  (LUVOIR)\cite{Peterson2017}, to a mid-infrared mission such as the Large Interferometer for Exoplanets (LIFE)\cite{Quanz2021}\, which will use nulling interferometry to constrain the bulk and atmospheric parameters of terrestrial exoplanets.

At this point in time, it is essential to determine the science requirements of such missions through theoretical studies. From simulated observations of spectra of exoplanets, it is possible to infer estimates on the bulk parameters and the atmospheric composition and structure of the exoplanets. This can be done through Bayesian retrieval routines, which apply Bayesian statistics to identify the best combination of physical parameters that could explain the data. This technique is still the gold standard in the exoplanet community for the interpretation of spectroscopic observations\cite{Madhusudhan_2018}. In the context of future missions, it is important to perform retrievals of simulated atmospheres in order to determine the required resolution (R), signal-to-noise ratio (S/N), and wavelength range to characterize the atmosphere of an exoplanet. This is particularly relevant when searching for life on other planets since living organisms can substantially modify the atmosphere of the planet they inhabit.

Bayesian frameworks are composed of two main routines: a radiative transfer model that can compute spectra based on a set of parameters; and a parameter estimation routine, which evaluates how well the spectrum that the radiative transfer model produces fits the data, then uses this inferred knowledge to iteratively update the set of parameters. Ultimately, we are left with ``posterior" distributions for the various parameters, that provide a statistical estimate of the set of parameters that best explains the data.

For the radiative transfer model to properly compute a spectrum, it is necessary to calculate the absorption and the emission of each layer of the atmosphere, which depends on the temperature and pressure of the layer, as well as their composition. This can be done during the computation of the spectrum. However, this slows down massively the retrieval run since millions of spectra are produced during the iterative process. For this reason, opacity tables can also be computed in advance at various pressures and temperatures and then queried by the retrieval model when needed. To compute opacity tables it is necessary to take into account the effects of pressure, temperature, and atmospheric composition at the various points of interest. Such effects cause a broadening of the line, which generally has a Voigt-shaped profile (a convolution between the pressure-broadened Lorentz profile and a temperature-broadened Gaussian profile). Then, to avoid the wings of the profile to be infinitely wide (an unrealistic scenario), a cut-off on the line wing must be applied. Opacity tables can change between the various radiative transfer models in use since there is no agreement in the community concerning the choice of the line list provenance, pressure broadening coefficients, and wing cut-offs. This can cause substantial differences in the computation of emission spectra\cite{2017ApJ...850..150B}, as well as downgrading the quality of the results in retrieval studies.

In our first retrieval paper published in the context of a paper series for the LIFE mission (LIFE Paper III\cite{konrad2021large}), we used the same input opacity tables to produce the simulated observation and to perform the retrievals. This led to a very precise characterization of the atmosphere. Then, in LIFE Paper V\cite{Alei2022}, we performed retrievals on simulated spectra produced by a separate radiative transfer model, as well as different input opacity tables\cite{Rugheimer2018}. Because of such differences, the retrieval output showed biases and offsets in the results. 
While differences between the various radiative transfer models exist and must be quantified\footnote{This effort requires a deep inter-model comparison that involves the whole community. An example of inter-model comparison is the BUFFET NExSS working group\cite{Fauchez2021}.}, it is important to assess how using separate input opacity tables in the retrievals can be a source of biases and systematics.
For this study, we performed the first analysis on the impact that the use of different input opacity tables has on retrievals. We focused on quantifying the effect of using different/newer line list databases, different broadening coefficients, and different line cut-off thresholds.

We describe the methodology and the retrieval runs that we consider in this study in Section~\ref{sec:methods}. We show the main results in Section~\ref{sec:results} and we discuss them in Section~\ref{sec:discussion}, also by putting them in context with other works. Finally, we summarize the main findings in Section~\ref{sec:conclusion}.

\section{METHODS}\label{sec:methods}

\begin{figure} [ht]
   \begin{center}
   \begin{tabular}{c} %% tabular useful for creating an array of images 
   \includegraphics[width=\textwidth]{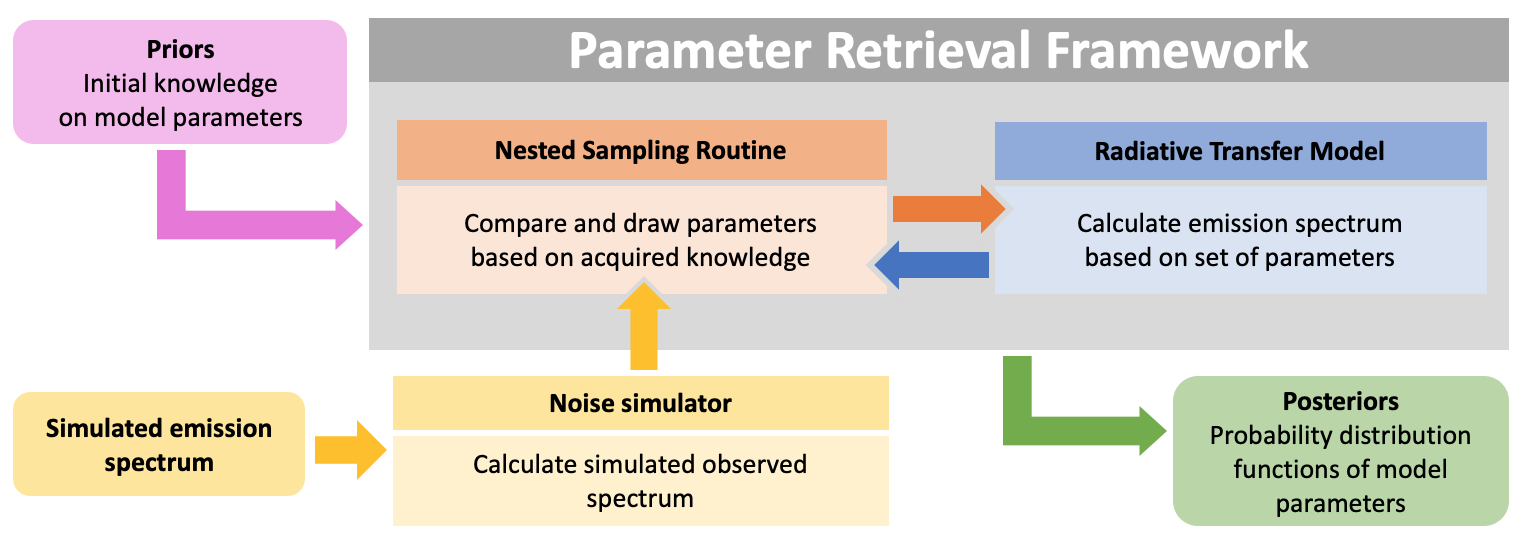}
   \end{tabular}
   \end{center}
   \caption[framework] 
%>>>> use \label inside caption to get Fig. number with \ref{}
   { \label{fig:flowchart} 
Schematic illustrating the retrieval framework (adapted from LIFE Paper III\cite{konrad2021large}).}
   \end{figure}

Similar to LIFE Paper III\cite{konrad2021large} and LIFE Paper V\cite{Alei2022}, we performed retrievals on a simulated input spectrum using the Bayesian retrieval routine that we developed. A qualitative flowchart of the retrieval process is shown in Figure~\ref{fig:flowchart}. We will describe the various elements of the retrieval framework in the remainder of this section.

   \subsection{Observed spectrum}

We assumed to observe a cloud-free modern Earth twin at a 1 AU orbit around a Sun-like star, 10 pc away from the observer. The input spectrum was published in Rugheimer \& Kaltenegger (2018)\cite{Rugheimer2018}. The authors used a 1D convective-radiative transfer code coupled with a 1D photochemistry model and a 1D climate model to produce a self-consistent pressure-temperature profile and altitude-dependent chemical abundances. These were then fed to a line-by-line radiative transfer model to calculate thermal spectra. This radiative transfer model made use of HITRAN~2016 \cite{Gordon2017} transitions, assumed air broadening, and a hard wing cut-off at 25~cm$\mathrm{^{-1}}$ from the core of each line. 
Since this spectrum was used also in LIFE Paper V as part of a larger set of retrieval runs, we refer the reader to our previous publication for further details on its computation. 

The simulated emission spectrum was then fed to the noise simulator for LIFE, LIFE\textsc{sim} (described in LIFE Paper II\cite{Dannert2022}). LIFE\textsc{sim} calculated the expected wavelength-dependent S/N for an observation obtained with LIFE of the target exoplanet, assuming all the major astrophysical noise sources (zodiacal and exo-zodiacal dust, stellar leakage) and the nominal architecture of the LIFE interferometer currently assumed (see Table~\ref{tab:lifesim}).

\begin{table}[]
\caption{LIFE\textsc{sim}: simulation parameters. See LIFE Paper I\cite{Quanz2021} and LIFE Paper II\cite{Dannert2022} for details. }              % title of Table
\label{tab:lifesim}      % is used to refer this table in the text
\centering                                      % used for centering table
\begin{tabular}{|l |l|}          % centered columns (4 columns)
                % inserts double horizontal lines
\hline
\textbf{Parameter} &\textbf{Value} \\    % table heading
\hline                                 % inserts single horizontal line
    Detector quantum efficiency & 0.7\\      % inserting body of the table
    \hline      
    Total instrument throughput & 0.05\\
    \hline      
    Minimum Wavelength & 4~$\mu$m       \\
    \hline      
    Maximum Wavelength & 18.5~$\mu$m   \\
    \hline      
    Spectral Resolution & 50      \\
    \hline      
    S/N at 11.2 $\mu$m  & 10\\
    \hline      
    Interferometric Baseline & 10-100~m \\
    \hline      
    Aperture Diameter & 2~m \\
    \hline      
    Exozodi level & 3 $\times$ local zodiacal dust \\
    \hline      
    Planet radius & 1~${R_\oplus}$\\
    \hline      
    Distance to the system & 10~pc\\

\hline
\end{tabular}

\end{table}

\subsection{Retrieval Framework}

Our retrieval framework (already validated and used in previous publications\cite{konrad2021large,Alei2022}) is composed of two main routines:
\begin{itemize}
    \item \texttt{petitRADTRANS}\cite{Molliere2019}, a 1D radiative transfer model that can calculate planetary spectra starting from a set of input parameters (planetary mass and radius, thermal profile, and chemical composition of the atmosphere). Spectra are calculated using correlated-k tables for the opacities which must be computed a priori. 
    \item \texttt{pyMultiNest}\cite{Buchner:PyMultinest}, which applies the Nested Sampling algorithm\cite{Skilling:Nested_Sampling} to evaluate the combinations of parameters that best fit the observed spectra. 
\end{itemize}

For more details about the Bayesian retrieval framework, we refer the reader to LIFE Paper III\cite{konrad2021large}.

\subsubsection{Priors}

To make our results consistent with the analysis performed on the same spectrum in LIFE Paper V\footnote{For more details, the retrieval results on the modern Earth cloud-free spectrum are labeled in LIFE Paper V with the ``MOD-CF" identifier.}, we kept the same prior space. The parameters that were considered in this study and their prior distributions are listed in Table~\ref{tab:priors}. We also report the expected values of each parameter as they were assumed in Rugheimer \& Kaltenegger (2018)\cite{Rugheimer2018} (or derived from the input data, as explained below). These will be used to compare the results with the input.

In our retrieval framework, we approximate the thermal (P-T) profile of the atmosphere with a fourth-degree polynomial. This translates into 6 total parameters to be retrieved to determine the shape of the P-T profile (the five polynomial coefficients $a_{i}$ with $i \in [0,4]$ and the ground pressure $P_0$). The expected values of the P-T parameters are generated by fitting the input P-T profile\cite{Rugheimer2018} with a fourth-degree polynomial. 
The retrieval assumes altitude-invariant chemical profiles: each species is equally abundant (in units of mass fraction) throughout the whole atmosphere. This is an approximation, especially for some species in Earth's atmosphere that strongly vary with altitude (e.g., \ce{H2O}, \ce{O3}). However, this is a necessary trade-off to allow the retrieval analysis to run in reasonable computing time. Since we expect that the deeper layers of the atmosphere would contribute the most to the thermal emission spectrum, by virtue of the higher atmospheric density, we calculated the expected values of the abundances considering the weighted mean by the pressure of the input abundance profiles.

Other parameters that were considered in the retrieval are the mass and the radius of the planet. The Gaussian priors that we assume for these two parameters are based on the expected radius estimate that we would get from the search phase of LIFE, as described in LIFE Paper II\cite{Dannert2022}. This would also produce an estimate of the mass based on statistical mass-radius relations\cite{Kipping:Forecaster} (see LIFE Paper III for details).

\begin{table}[ht]
\caption{Summary of the parameters used in the retrievals, their expected values and their prior distributions (adapted from LIFE Paper V). $\mathcal{U}(x,y)$ represents a boxcar prior with a lower threshold $x$ and upper threshold $y$; $\mathcal{G}(\mu,\sigma)$ denotes a Gaussian prior with mean $\mu$ and standard deviation $\sigma$. For $a_4$ we choose a prior on $\sqrt[4]{a_4}$, and then take the fourth power to obtain $a_4$, to sample the small values more densely. }
\label{tab:priors}
\begin{center}       
\begin{tabular}{|l|l|l|l|} 
\hline
\textbf{Parameter} & \textbf{Description} & \textbf{Prior }& \textbf{Expected value}\\
\hline
$\sqrt[4]{a_4}$ &P-T Parameter (Degree 4)     & $\mathcal{U}(0.5,1.8)$ &0.694 \\
\hline
$a_3$           &P-T parameter (Degree 3)  &  $\mathcal{U}(0,100)$  & 18.644 \\
\hline
$a_2$           &P-T Parameter (Degree 2)  &  $\mathcal{U}(0,500)$   & 110.331 \\
\hline
$a_1$           &P-T Parameter  (Degree 1) &  $\mathcal{U}(0,500)$   & 177.997 \\
\hline
$a_0$           &P-T Parameter (Degree 0) &  $\mathcal{U}(0,1000)$  & 293.60\\
\hline

$\log_{10}\left(P_0\left[\mathrm{bar}\right]\right)$&Surface Pressure    & $\mathcal{U}(-4,3)$& 0.007\\
\hline
$R_{\text{pl}}\,\left[R_\oplus\right]$&Planet Radius (bulk value)&  $\mathcal{G}(1.0,0.2)$ & 1.000\\ 
\hline
$\log_{10}\left(M_{\text{pl}}\,\left[M_\oplus\right]\right)$&Planet Mass (bulk value) & $\mathcal{G}(0.0,0.4)$& 0.000\\
\hline
$\log_{10}(\mathrm{N_2})$    &\ce{N2}  Mass Fraction  & $\mathcal{U}(-15,0)$  & -0.113 \\
\hline
$\log_{10}(\mathrm{O_2})$    & \ce{O2} Mass Fraction     & $\mathcal{U}(-15,0)$  & -0.631\\
\hline
$\log_{10}(\mathrm{H_2O})$   & \ce{H2O} Mass Fraction           & $\mathcal{U}(-15,0)$ & -2.607\\
\hline
$\log_{10}(\mathrm{CO_2})$   & \ce{CO2} Mass Fraction     & $\mathcal{U}(-15,0)$ & -3.265 \\
\hline
$\log_{10}(\mathrm{CH_4})$   & \ce{CH4} Mass Fraction         & $\mathcal{U}(-15,0)$ & -6.028 \\
\hline
$\log_{10}(\mathrm{O_3})$    & \ce{O3} Mass Fraction            & $\mathcal{U}(-15,0)$ & -6.436 \\
\hline
$\log_{10}(\mathrm{CO})$     &\ce{CO}  Mass Fraction       & $\mathcal{U}(-15,0)$& -7.215 \\
\hline
$\log_{10}(\mathrm{N_2O})$   &\ce{N2O} Mass Fraction     & $\mathcal{U}(-15,0)$& -6.343\\
\hline 
\end{tabular}
\end{center}
\end{table}

\subsubsection{Opacity tables}

To make this study possible, we computed additional correlated-k tables assuming various line list databases (most importantly HITRAN~2020\cite{GORDON2022107949}, the latest version of the HITRAN database to date). We also considered various line wing cut-offs in the computation (namely 25 and 100~cm$\mathrm{^{-1}}$, two values commonly used in the community). We used the opacity calculator that was used to precompute the default \texttt{petitRADTRANS} line opacities, on a pressure-temperature grid that covers the space between $10^{-6}$ and $10^3$ bar, and between 200 and 3000 K. For \ce{CH4} and \ce{O3}, the new tables have been calculated only up to 2500 K and 1000 K respectively, due to the lack of partition functions values for these molecules above the threshold temperature. This shouldn't be an issue in the current scenario since our retrievals are limited to temperate planets. The line opacities were calculated from 110 nm to 250 $\mu$m and at a spacing of $\lambda/\Delta\lambda=10^{6}$. These were then converted into correlated-k tables at $\lambda/\Delta\lambda=10^{3}$ and subsequently binned down to $R=50$, the spectral resolution considered in this study. For more details about the computation of the absorption cross-sections and the correlated-k opacity tables, we refer the reader to Molli\`ere et al. (2015)\cite{Molliere2015}, Molli\`ere et al. (2019)\cite{Molliere2019} and references therein.
The new tables can now be used in addition to the set of correlated-k tables provided on the \texttt{petitRADTRANS} website\footnote{\url{https://petitradtrans.readthedocs.io/en/latest/}}, as well as the \texttt{petitRADTRANS}-compatible tables that are available in ExoMolOP\cite{Chubb2021}.

\subsection{Runs}

We performed four different retrievals on the same cloud-free modern Earth spectrum as it would be observed with LIFE. The observed spectrum, the number of parameters, and their priors were kept fixed, while we assumed different opacity tables for the various species. Table~\ref{tab:setups} describes the details of the tables used in each run. We can describe the runs as follows:
\begin{itemize}
\item \emph{Run~1}. Heterogeneous opacities from various sources and assuming various broadening coefficients and cut-offs. These opacity tables are part of the default \texttt{petitRADTRANS} opacity folder. This retrieval run is taken from LIFE Paper V and used for comparison.
\item \emph{Run~2}. For most species, \texttt{petitRADTRANS}-compatible opacity tables provided by ExoMolOP\cite{Chubb2021}, assuming \ce{H}-\ce{He} broadening and cut-off from Eq. 6 from the ExoMolOP paper\cite{Chubb2021}. Opacity table for \ce{O3} from HITRAN~2012, assuming air broadening and sub-Lorentzian cut-off from Hartmann et al. (2002)\cite{Hartmann2002}.
\item \emph{Run~3}. Opacity tables calculated from HITRAN~2020\cite{GORDON2022107949} assuming air broadening and wing cut-off at 100~$\mathrm{cm^{-1}}$ from the line core.
\item \emph{Run~4}. Opacity tables calculated from HITRAN~2020\cite{GORDON2022107949} assuming air broadening and wing cut-off at 25~$\mathrm{cm^{-1}}$ from the line core.

\end{itemize}

For every retrieval run, we consider the same set of continuum-induced absorption opacity tables (\ce{CO2}-\ce{CO2}, \ce{N2}-\ce{O2}, \ce{O2}-\ce{O2}, \ce{N2}-\ce{N2}) which were used in LIFE Paper III\cite{konrad2021large} and LIFE Paper V\cite{Alei2022}).
We also include Rayleigh scattering of \ce{O2}, \ce{CO}, \ce{CH4}, \ce{CO2}, \ce{N2}, and \ce{H2O}. In the calculation, we assume a surface reflectance of 0.1 over all wavelengths and we calculate the thermal atmospheric scattering through Feautrier's method\footnote{We refer to LIFE Paper V\cite{Alei2022} for more details about the scattering treatment.}.

We point out that neither of these retrieval runs uses the same opacity tables that Rugheimer \& Kaltenegger (2018)\cite{Rugheimer2018} used to produce their spectrum. We, therefore, expect to find discrepancies and systematics. However, we can still assess which retrieval setup(s) would allow us to retrieve better results. 

\begin{landscape}
\begin{table}[htbp]
\renewcommand{\arraystretch}{1.5}

\caption{Details of the opacity tables used in the four runs. The runs are associated with a specific color, which will be used in all the figures.} 
\label{tab:setups}
\begin{center}       
\begin{tabular}{|l|l|l|l|l|} 
\hline
\rule[-1ex]{0pt}{3.5ex}\textbf{ Run} & \textbf{Species} & \textbf{Line database }& \textbf{Broadening }& \textbf{Wing cut-off }\\
\hline

\multirow{6}{*}{\large{1} \crule[blue]{0.55cm}{0.35cm}} & \ce{CO2} & HITEMP~2010 \cite{ROTHMAN20102139}  & $\gamma_{\mathrm{air}}$ & Burch (1969)\cite{Burch:69}\\\cline{2-5}
                   & \ce{O3} & HITRAN~2012\cite{Rothman2013}  & $\gamma_{\mathrm{air}}$ & Hartmann et al. (2002)\cite{Hartmann2002}\\\cline{2-5}
                   & \ce{CH4} & ExoMol~2014\cite{Yurchenko2014} &  Sharp \& Burrows (2007, Eq. 15)\cite{Sharp2007}  & Hartmann et al. (2002)\cite{Hartmann2002}\\\cline{2-5}
                   & \ce{CO} & HITEMP~2010 \cite{ROTHMAN20102139}  &  $\gamma_{\mathrm{air}}$& Hartmann et al. (2002)\cite{Hartmann2002}\\\cline{2-5}
                   & \ce{H2O} & HITEMP~2010 \cite{ROTHMAN20102139}  &  $\gamma_{\mathrm{air}}$& Hartmann et al. (2002)\cite{Hartmann2002}\\\cline{2-5}
                   & \ce{N2O} &  ExoMol~2021\cite{Chubb2021}& $\gamma_{\ce{H}},\ \gamma_{\ce{He}}$ & ExoMol~2021 (Eq. 6)\cite{Chubb2021}  \\\hline
\hline
\multirow{6}{*}{\large{2} \crule[pink]{0.55cm}{0.35cm}} & \ce{CO2} & ExoMol~2021\cite{Chubb2021} &  $\gamma_{\ce{H}},\ \gamma_{\ce{He}}$ & ExoMol~2021 (Eq. 6)\cite{Chubb2021}  \\\cline{2-5}
                   & \ce{O3} & HITRAN~2012\cite{Rothman2013}  & $\gamma_{\mathrm{air}}$ & Hartmann et al. (2002)\cite{Hartmann2002}\\\cline{2-5}
                   & \ce{CH4} & ExoMol~2021\cite{Chubb2021} &   $\gamma_{\ce{H}},\ \gamma_{\ce{He}}$& ExoMol~2021 (Eq. 6)\cite{Chubb2021} \\\cline{2-5}
                   & \ce{CO} & ExoMol~2021\cite{Chubb2021} &   $\gamma_{\ce{H}},\ \gamma_{\ce{He}}$& ExoMol~2021 (Eq. 6)\cite{Chubb2021}  \\\cline{2-5}
                   & \ce{H2O} & ExoMol~2021\cite{Chubb2021} &   $\gamma_{\ce{H}},\ \gamma_{\ce{He}}$& ExoMol~2021 (Eq. 6)\cite{Chubb2021}  \\\cline{2-5}
                   & \ce{N2O} & ExoMol~2021\cite{Chubb2021} &   $\gamma_{\ce{H}},\ \gamma_{\ce{He}}$& ExoMol~2021 (Eq. 6)\cite{Chubb2021} \\\hline    
\hline                  
\multirow{6}{*}{\large{3} \crule[orange]{0.55cm}{0.35cm}} & \ce{CO2} & HITRAN~2020\cite{GORDON2022107949} & $\gamma_{\mathrm{air}}$ & 100~$\mathrm{cm^{-1}}$\\\cline{2-5}
                   & \ce{O3} &  HITRAN~2020\cite{GORDON2022107949}  & $\gamma_{\mathrm{air}}$ & 100~$\mathrm{cm^{-1}}$\\\cline{2-5}
                   & \ce{CH4} & HITRAN~2020\cite{GORDON2022107949}   & $\gamma_{\mathrm{air}}$ & 100~$\mathrm{cm^{-1}}$\\\cline{2-5}
                   & \ce{CO} &  HITRAN~2020\cite{GORDON2022107949}  &  $\gamma_{\mathrm{air}}$& 100~$\mathrm{cm^{-1}}$\\\cline{2-5}
                   & \ce{H2O} & HITRAN~2020\cite{GORDON2022107949}  & $\gamma_{\mathrm{air}}$ & 100~$\mathrm{cm^{-1}}$\\\cline{2-5}
                   & \ce{N2O} &  HITRAN~2020\cite{GORDON2022107949}  & $\gamma_{\mathrm{air}}$ & 100~$\mathrm{cm^{-1}}$\\  \hline  
\hline                   
\multirow{6}{*}{\large{4} \crule[green]{0.55cm}{0.35cm}} & \ce{CO2} &  HITRAN~2020\cite{GORDON2022107949}  &  $\gamma_{\mathrm{air}}$&  25~$\mathrm{cm^{-1}}$ \\\cline{2-5}
                   & \ce{O3} & HITRAN~2020\cite{GORDON2022107949}   &  $\gamma_{\mathrm{air}}$&  25~$\mathrm{cm^{-1}}$ \\\cline{2-5}
                   & \ce{CH4} & HITRAN~2020\cite{GORDON2022107949}   & $\gamma_{\mathrm{air}}$ & 25~$\mathrm{cm^{-1}}$  \\\cline{2-5}
                   & \ce{CO} &  HITRAN~2020\cite{GORDON2022107949}  & $\gamma_{\mathrm{air}}$ &  25~$\mathrm{cm^{-1}}$ \\\cline{2-5}
                   & \ce{H2O} &  HITRAN~2020\cite{GORDON2022107949}  & $\gamma_{\mathrm{air}}$ &  25~$\mathrm{cm^{-1}}$ \\\cline{2-5}
                   & \ce{N2O} & HITRAN~2020\cite{GORDON2022107949}   & $\gamma_{\mathrm{air}}$ &  25~$\mathrm{cm^{-1}}$ \\\hline
                   
  \hline
\end{tabular}
\end{center}
\end{table}

\end{landscape}

\section{RESULTS}\label{sec:results}

In this section, we will describe the main outputs of the four retrieval runs described in Table~\ref{tab:setups}. We start by comparing the retrieved emission spectra in Section~\ref{sec:spectra} and the pressure-temperature profiles in Section~\ref{sec:pt}. Finally, we will focus on the posterior distributions for the bulk parameters and the atmospheric composition in Section~\ref{sec:posteriors}.

\subsection{Retrieved Emission Spectra}\label{sec:spectra}

The retrieved spectra are directly calculated from the posterior distributions, which are the direct output of the retrieval run. Given every set of parameters in the posterior space, it is possible to produce a spectrum.

\begin{figure}[ht]
\centering
\subfloat{\includegraphics[width=0.49\textwidth]{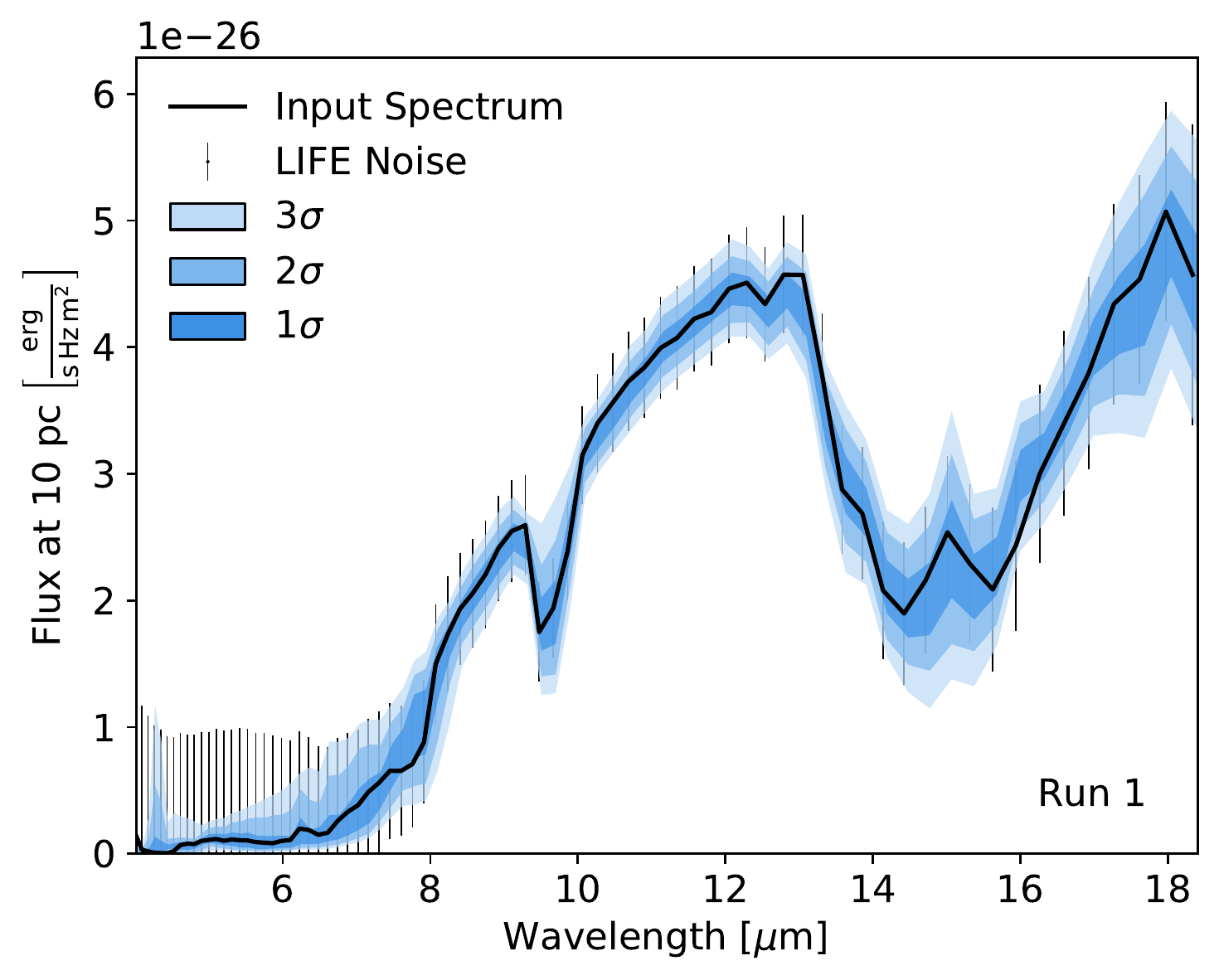}}
\,
\subfloat{\includegraphics[width=0.49\textwidth]{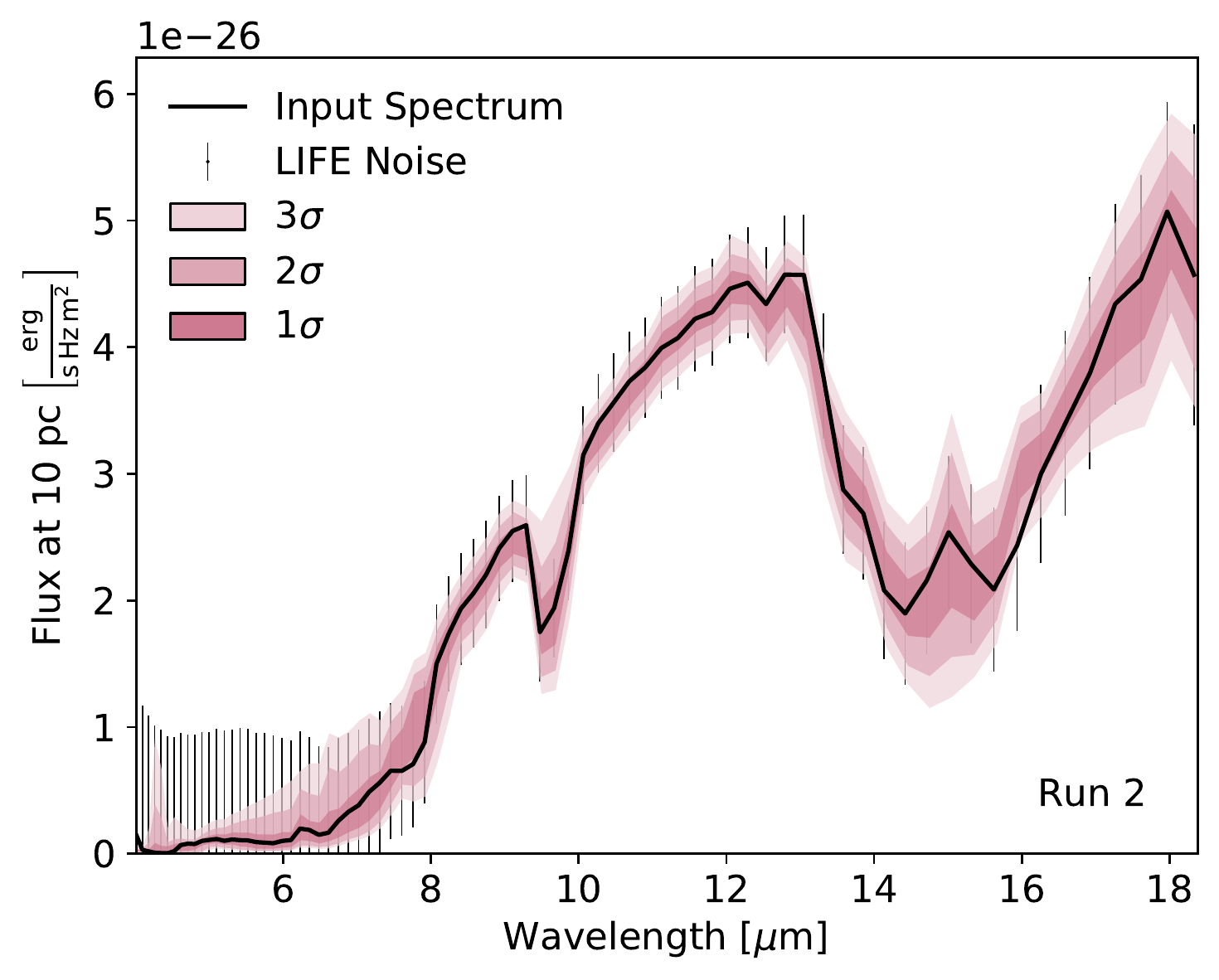}}
\\

\subfloat{\includegraphics[width=0.49\textwidth]{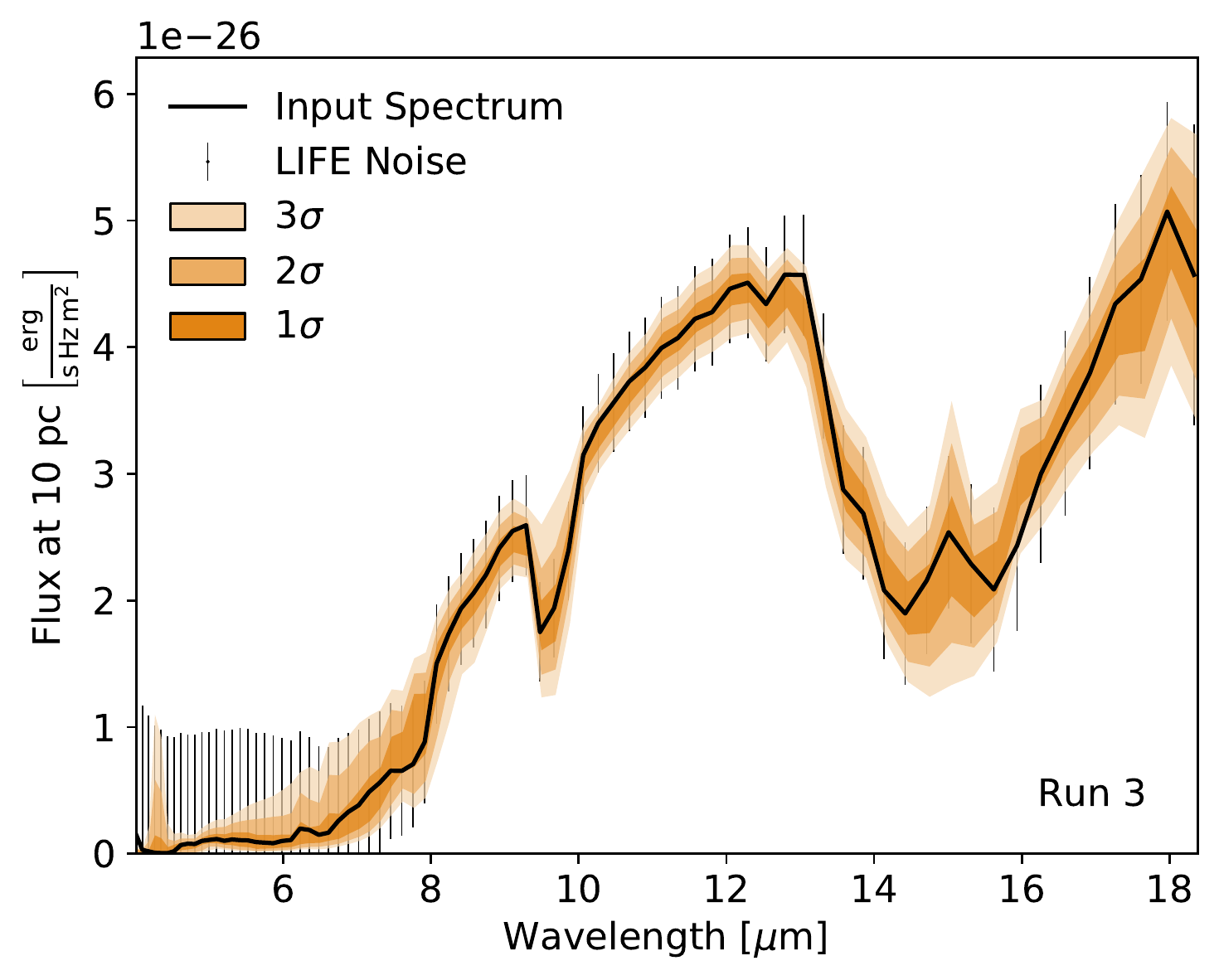}}
\,
\subfloat{\includegraphics[width=0.49\textwidth]{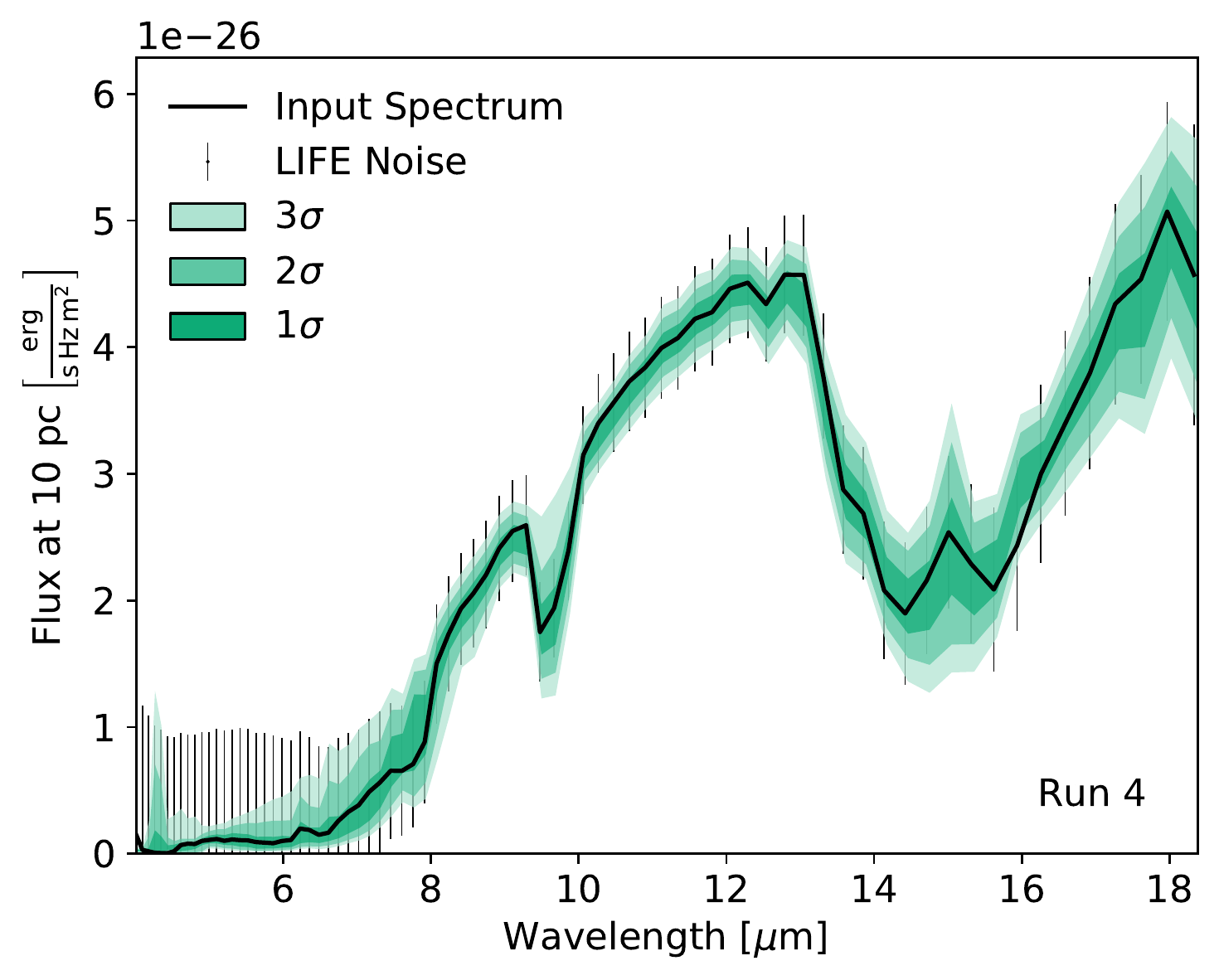}}
\,
 \caption[Spectra] 
%>>>> use \label inside caption to get Fig. number with \ref{}
   { \label{fig:spectra} 
Retrieved spectra compared to the input spectrum (black line) for the various runs. The gray error bars indicate the LIFE\textsc{sim} uncertainty. The color-shaded areas represent the confidence envelopes
(darker shading corresponds to higher confidence).}

\end{figure}

\begin{figure}[ht]
\centering
\subfloat{\includegraphics[width=\textwidth]{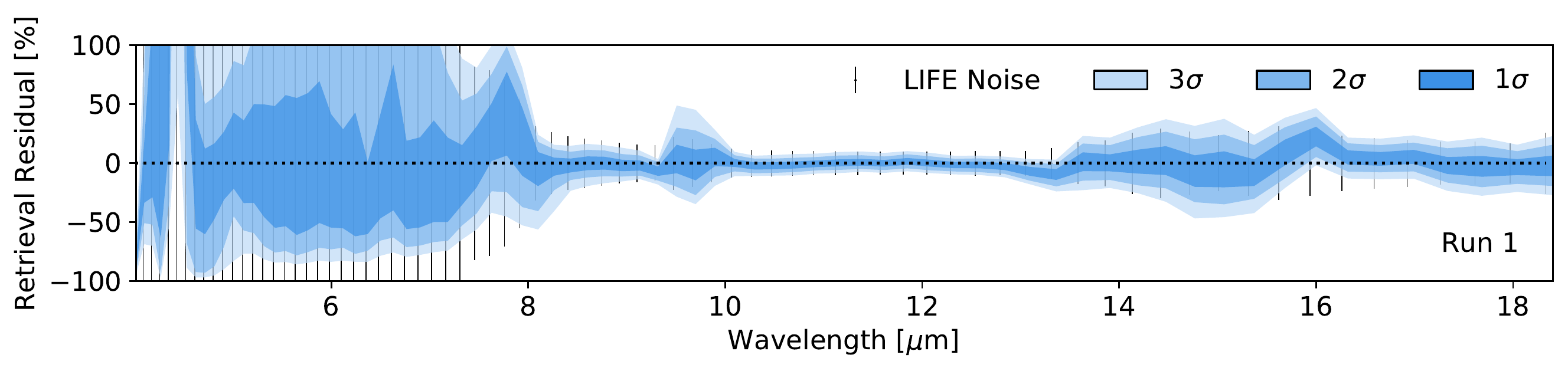}}
\\
\subfloat{\includegraphics[width=\textwidth]{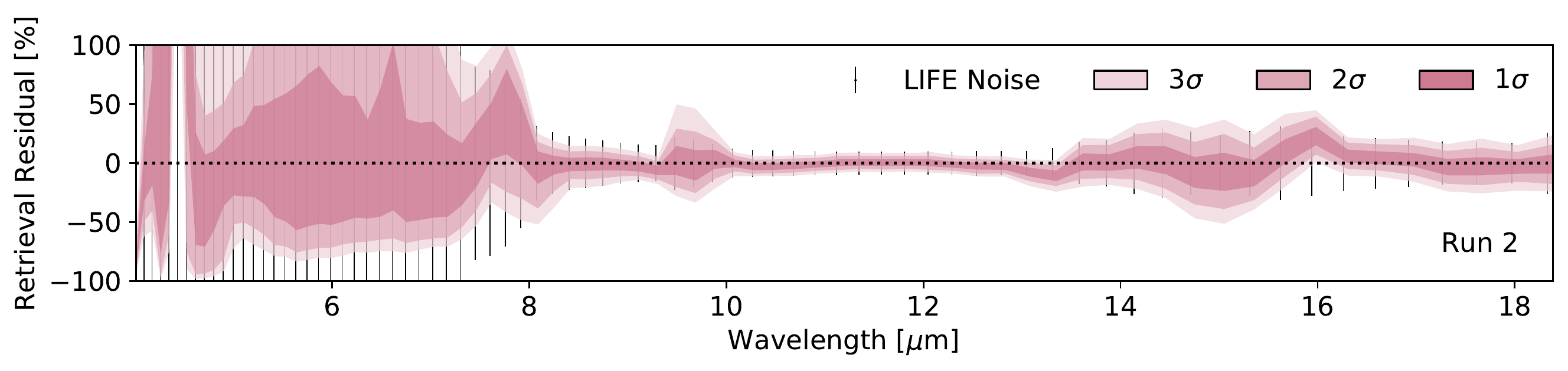}}
\\

\subfloat{\includegraphics[width=\textwidth]{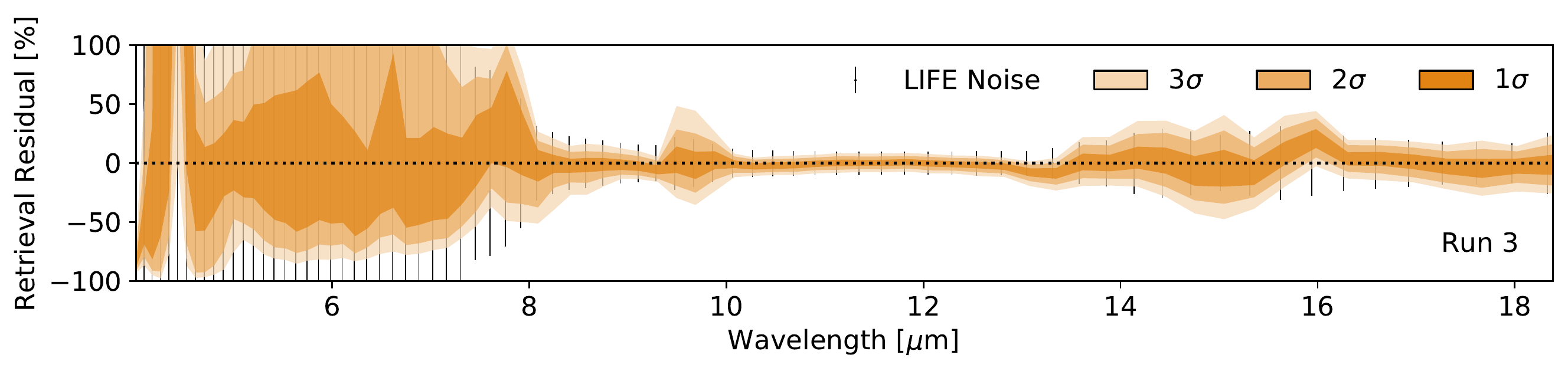}}
\\
\subfloat{\includegraphics[width=\textwidth]{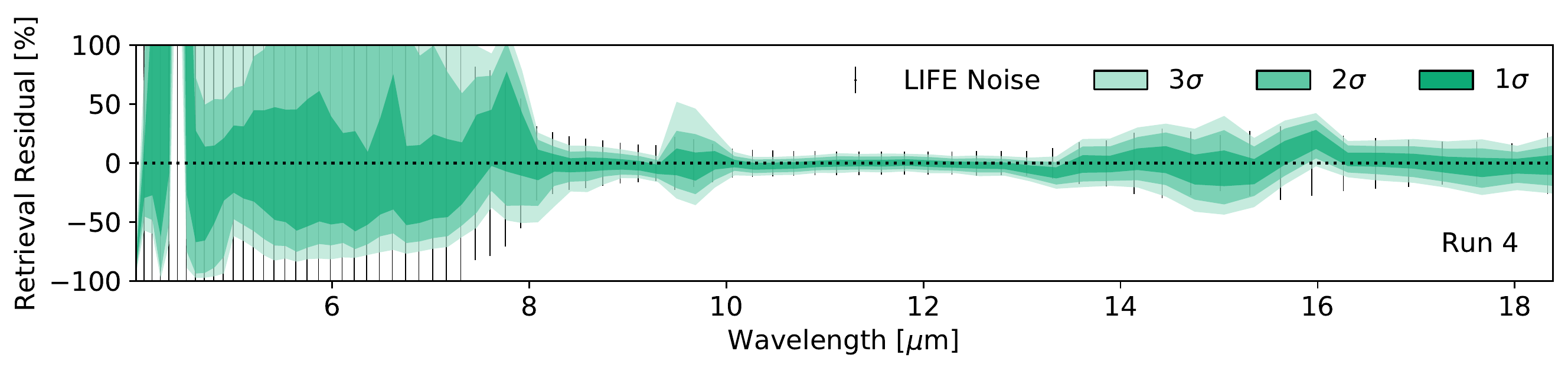}}
\,
 \caption[Spectra] 
%>>>> use \label inside caption to get Fig. number with \ref{}
   { \label{fig:residuals} 
Ratios between the retrieved flux and the input flux (in percent) for the various runs. The gray error bars indicate the LIFE\textsc{sim} uncertainty. The color-shaded areas represent the confidence envelopes (darker shading corresponds to higher confidence).}

\end{figure}

In Figure~\ref{fig:spectra}, we show the retrieved spectra for the four different runs, compared to the simulated observed spectrum, represented by a black solid line. The black error bars represent the noise level at each spectral point, as determined by LIFE\textsc{sim}. The color-shaded areas correspond to the confidence envelopes of the retrieved spectrum (in units of $\sigma$): the darker the region, the higher the confidence interval.

 From Figure~\ref{fig:spectra}, it is clear that every combination of parameters in the posterior space generates a spectrum that is, overall, well within the uncertainty of the observation (black error bars). The large majority of the retrieved spectra are consistent with the input spectrum, aside from some larger deviations between 8-10 $\mu$m and 14-16 $\mu$m (see the residuals in Figure~\ref{fig:residuals}, defined as the ratio between each retrieved flux point and the corresponding input flux point). Such differences are likely caused by the differences in the opacity tables used in the retrieval runs compared to the ones used by Rugheimer \& Kaltenegger (2018)\cite{Rugheimer2018}, probably also in terms of isotopic transitions. The opacity tables we computed only take into account the main isotope for all species; on the other hand, Rugheimer \& Kaltenegger might have considered line transitions of secondary isotopes in their opacities.
 
The main goal of a retrieval framework is to find the best subset of parameters to fit the observation. A retrieved spectrum close to the observed one only serves to show that the parameter estimation routine is working the way it was designed to do. This does not necessarily mean that each parameter has been correctly retrieved. Actually, it is common in atmospheric studies to have correlations and degeneracies between various parameters, which would allow us to produce similar emission spectra even with very different sets of parameters. For this reason, it is necessary to deepen our study into the analysis of the parameter space.

\subsection{P-T profiles}\label{sec:pt}

\begin{figure}[ht]
\centering
\subfloat{\includegraphics[width=.49\textwidth]{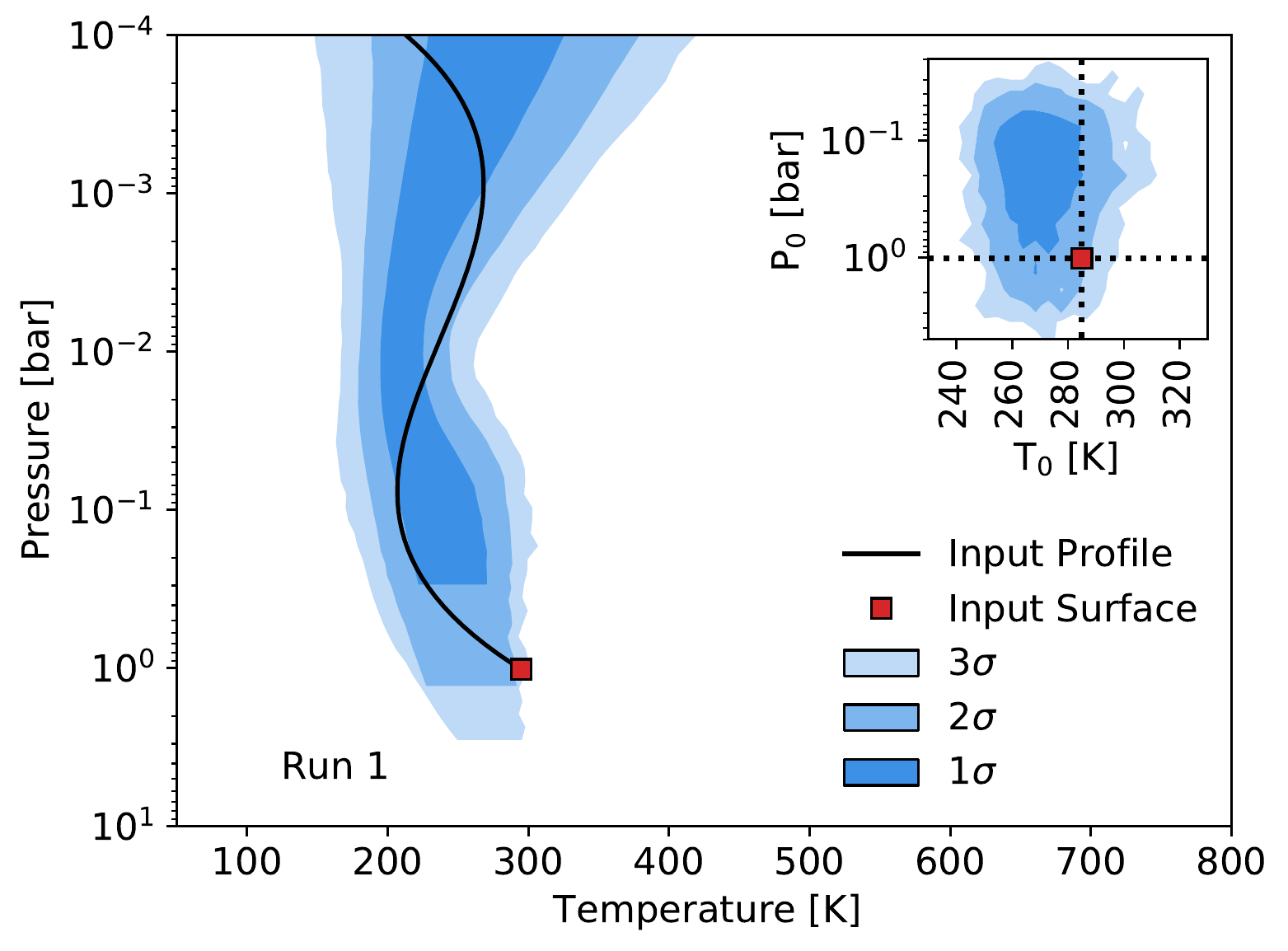}}
\,
\subfloat{\includegraphics[width=.49\textwidth]{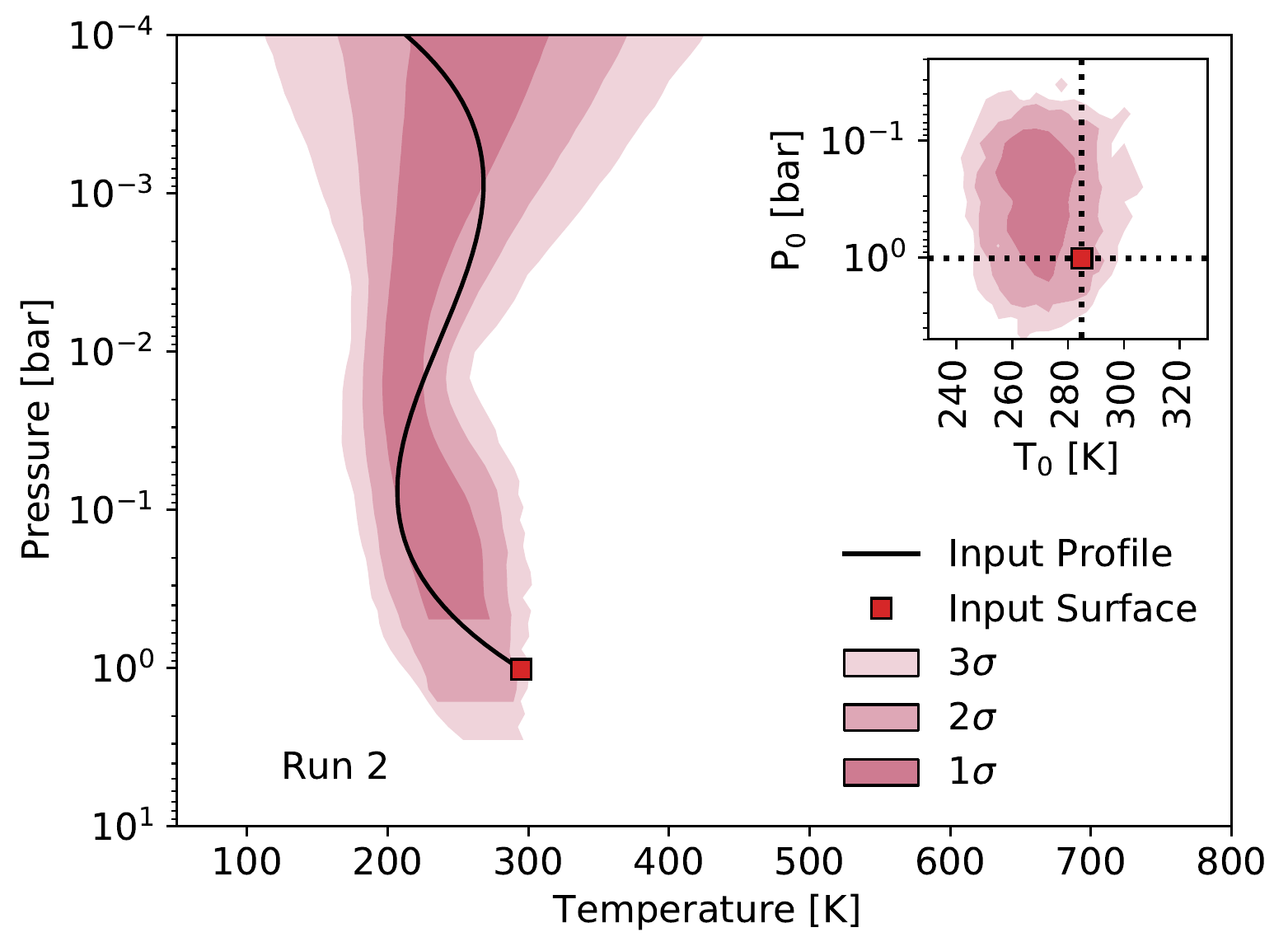}}
\\

\subfloat{\includegraphics[width=.49\textwidth]{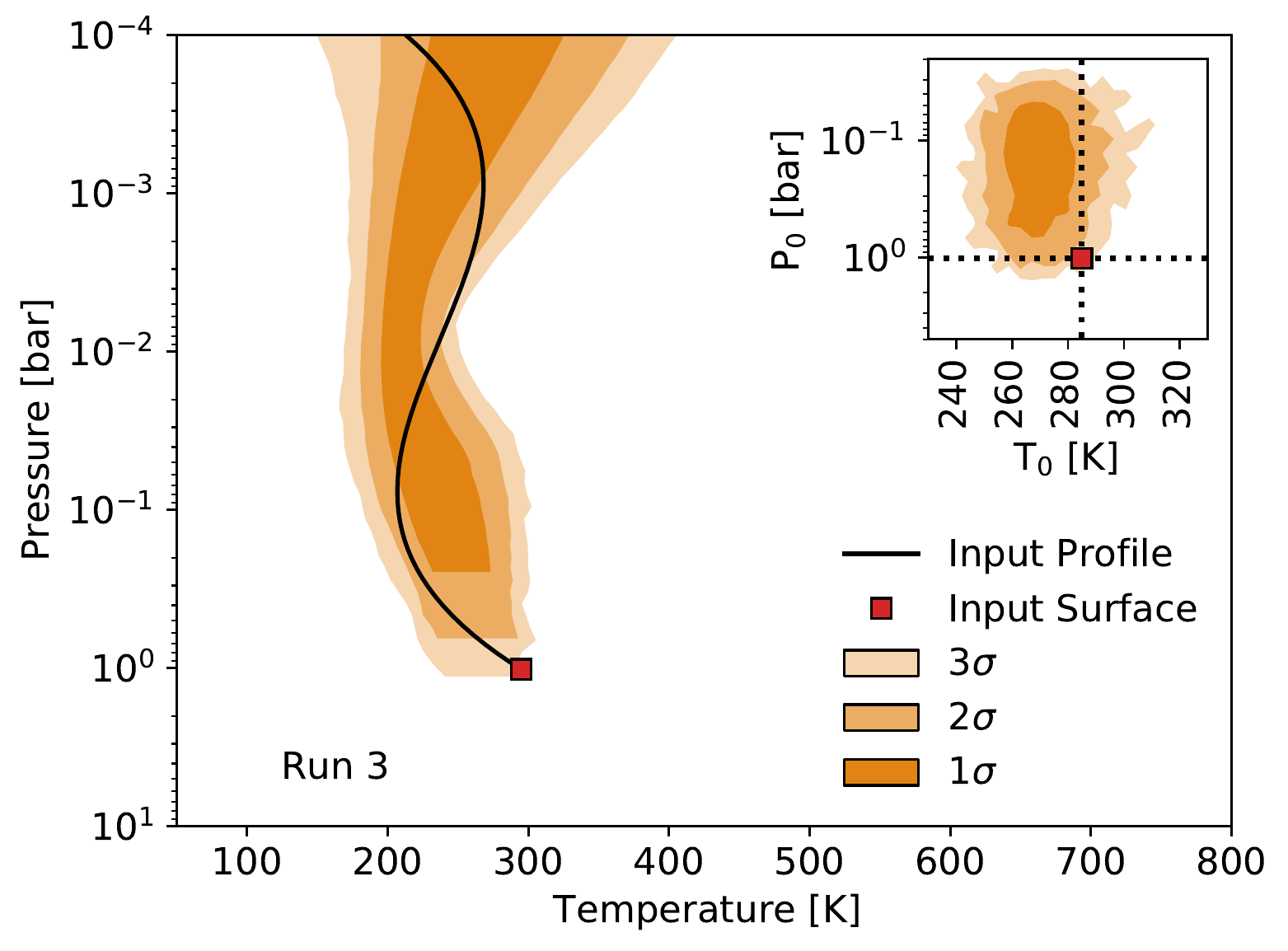}}
\,
\subfloat{\includegraphics[width=.49\textwidth]{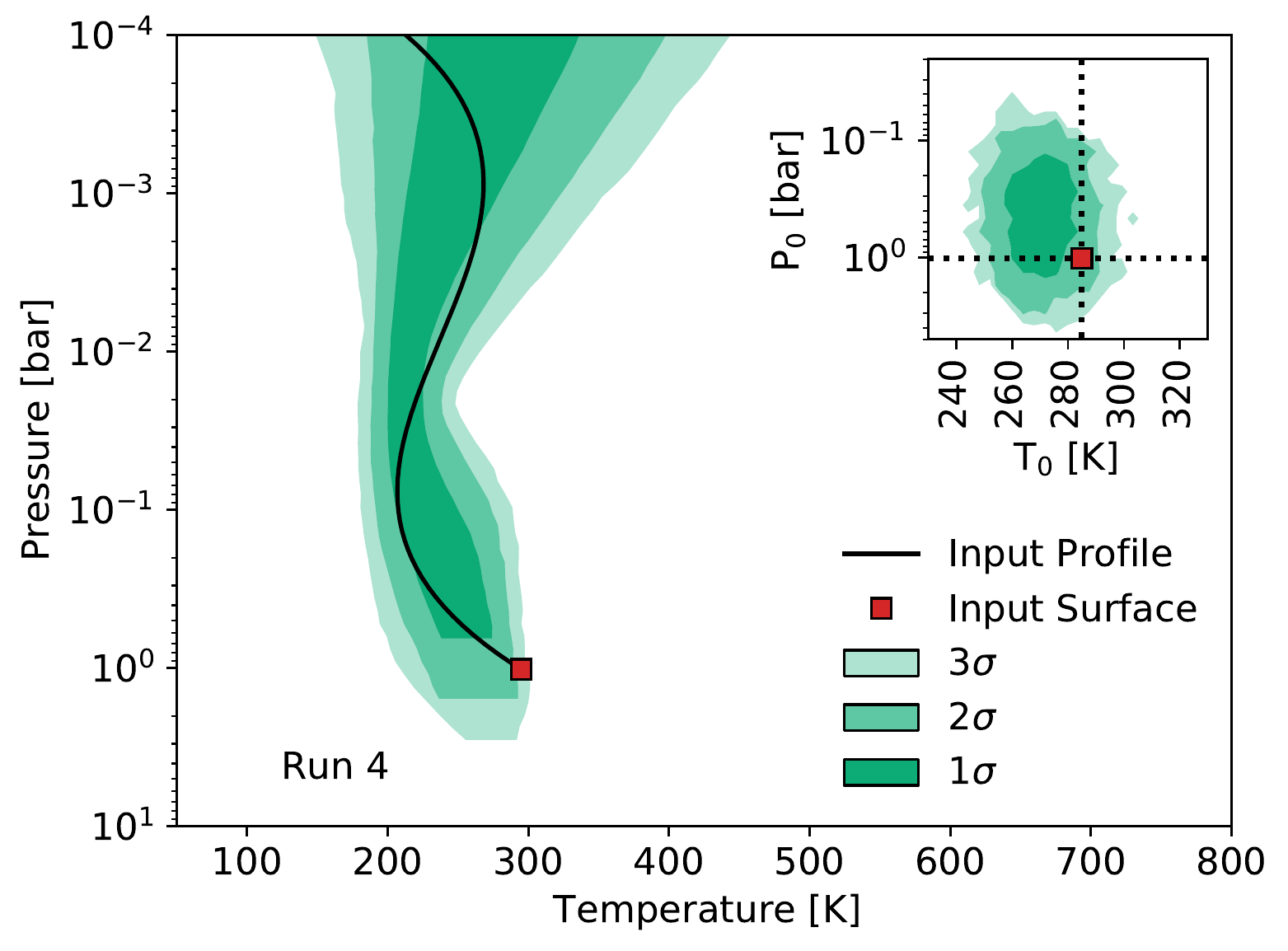}}
\,

 \caption[PT structure] 
%>>>> use \label inside caption to get Fig. number with \ref{}
   { \label{fig:ptstructure} 
Retrieved P-T profiles compared to the input profiles (black line) for the various runs. The red square markers represent the pressure-temperature point at the surface layer. The color-shaded areas represent the confidence envelopes
(darker shading corresponds to higher confidence).}
\end{figure}

In Figure~\ref{fig:ptstructure} we represent the retrieved pressure-temperature (P-T) profiles for every run. In black, we show the input P-T profile as calculated by Rugheimer \& Kaltenegger (2018)\cite{Rugheimer2018}. The inset plot shows a 2D histogram of the surface pressure $P_0$ versus the surface temperature $T_0$ posteriors, compared to the true values (red square marker).   We show the 3-$\sigma$, 2-$\sigma$, and 1-$\sigma$ confidence levels for the retrieved profile and the posterior distribution of the surface values in progressively darker shading.

For all the retrieval runs, there is a good agreement between the retrieved profile and the input in the deeper atmosphere and up to $\sim10^{-2}$ bar. in this region, the input profile lies generally within the 2-$\sigma$ envelope of the retrieved profile. At lower pressures, the uncertainty on the retrieved profile grows larger: this is expected since most of the features of the spectrum are determined by the lower and denser layers of the atmosphere. 

A general trend can be observed in all runs: the surface pressure and temperature are underestimated, with the retrieved estimates (within the 1-$\sigma$ uncertainty) being up to 10\% smaller than the input values. Specifically, this translates into an estimate of the surface pressure around a few tenths of a bar, compared to the standard 1 bar surface pressure value, and a temperature up to $\sim$ 25~K cooler than the input surface temperature. While the temperature discrepancy does not change with the different runs, a shift of the 1-$\sigma$ envelope of pressure posterior towards the input value is visible in Runs~2 and 4. The latter also shows a smaller variance in the ground pressure posterior compared to the other retrieval runs. The retrieval of the ground pressure is strictly correlated with the retrieval of the main absorbers in the atmosphere: this is the so-called pressure-abundance degeneracy, which we introduced in our previous works and which will be further described in the next subsection.

\subsection{Posteriors}\label{sec:posteriors}

\begin{figure}[ht]
\centering
\subfloat{\includegraphics[height=6cm]{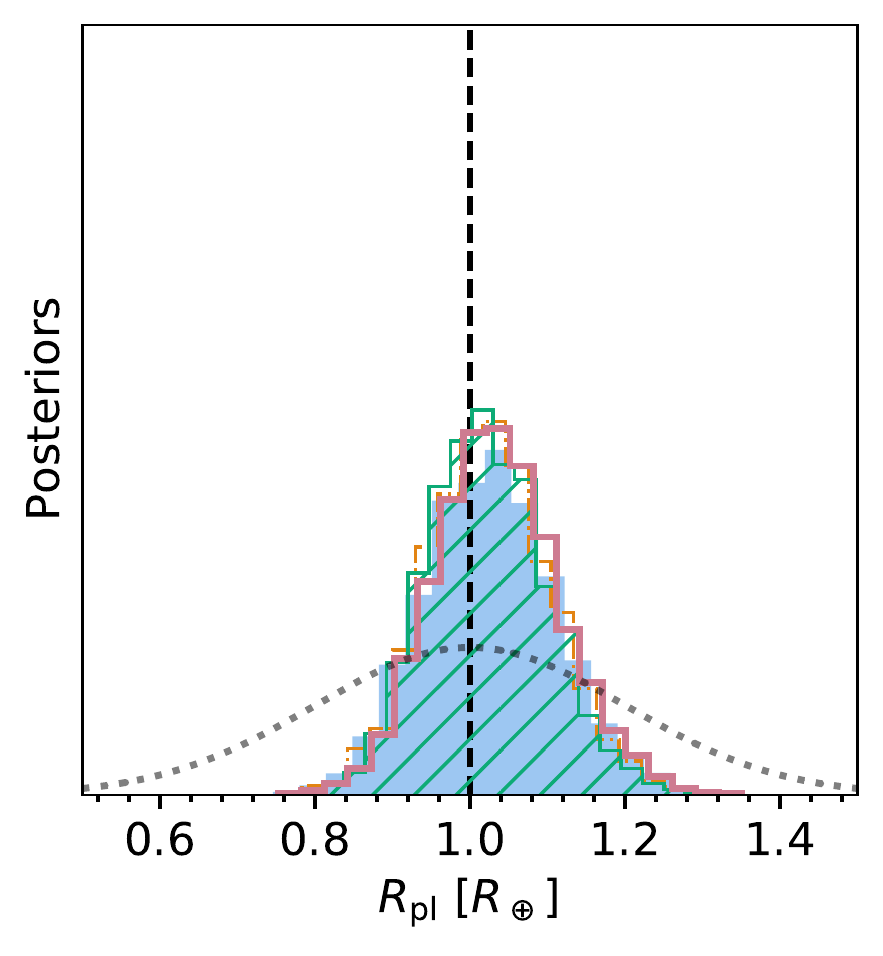}}
\,
\subfloat{\includegraphics[height=6cm]{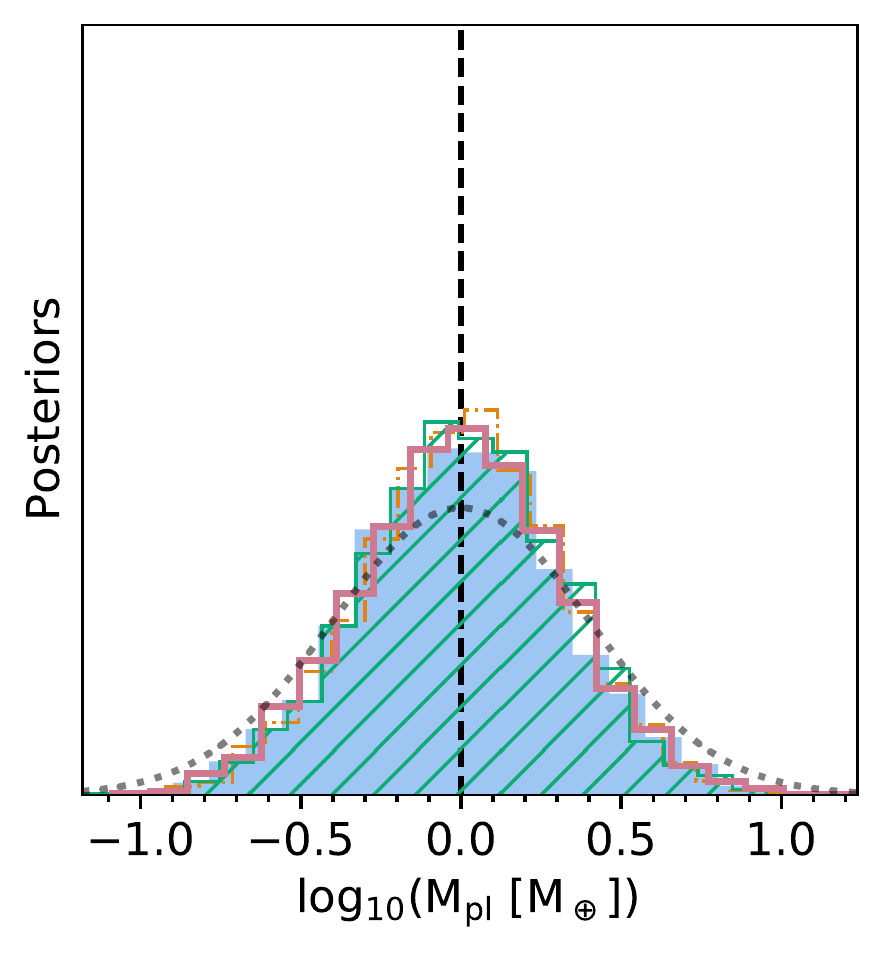}}
\\
\subfloat{\includegraphics[height=6cm]{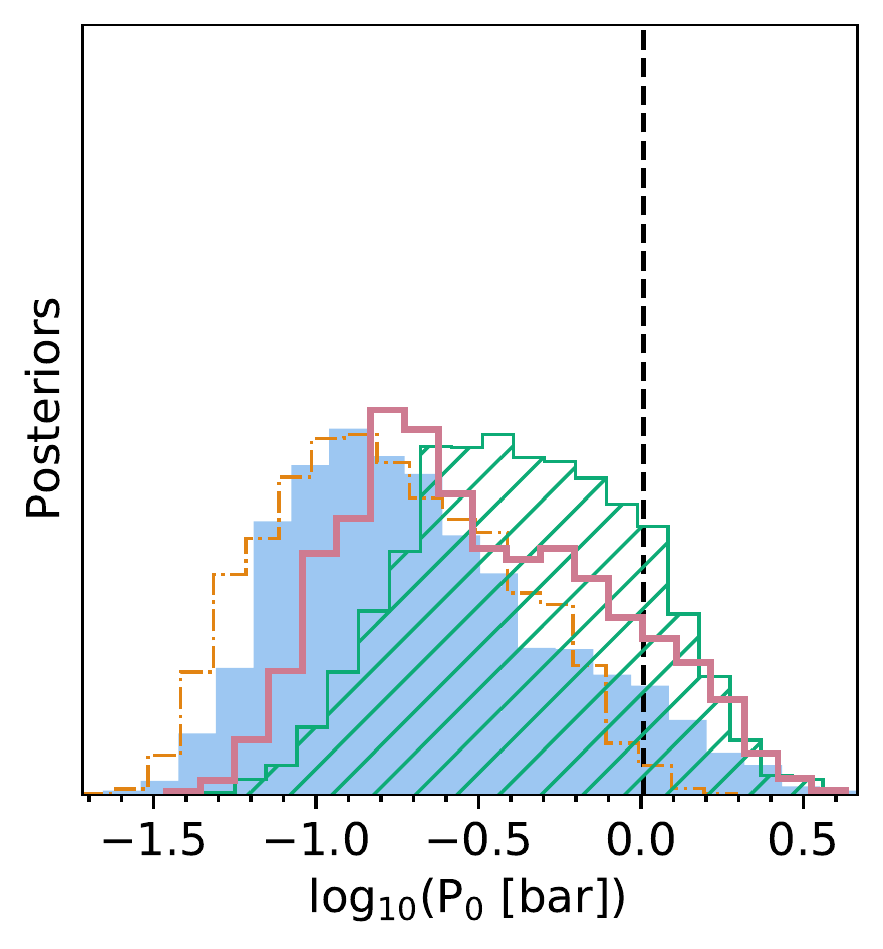}}
\,
\subfloat{\includegraphics[height=5.5cm]{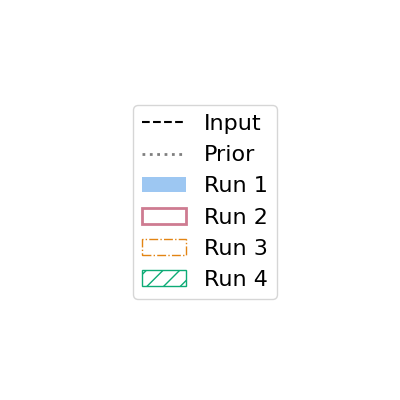}}
\\

 \caption[Posteriors] 
%>>>> use \label inside caption to get Fig. number with \ref{}
   { \label{fig:posteriors_bulk} 
 Posterior density distributions for the retrieved bulk parameters (planet radius, planet mass, ground pressure) for the four runs. The vertical, dashed lines mark the expected values for each parameter. For $R_{pl}$ and $\mathrm{log_{10}(M_{pl})}$, dotted lines show the corresponding priors for comparison.}

\end{figure}

\begin{figure}[ht]
\centering
\subfloat{\includegraphics[height=4.5cm]{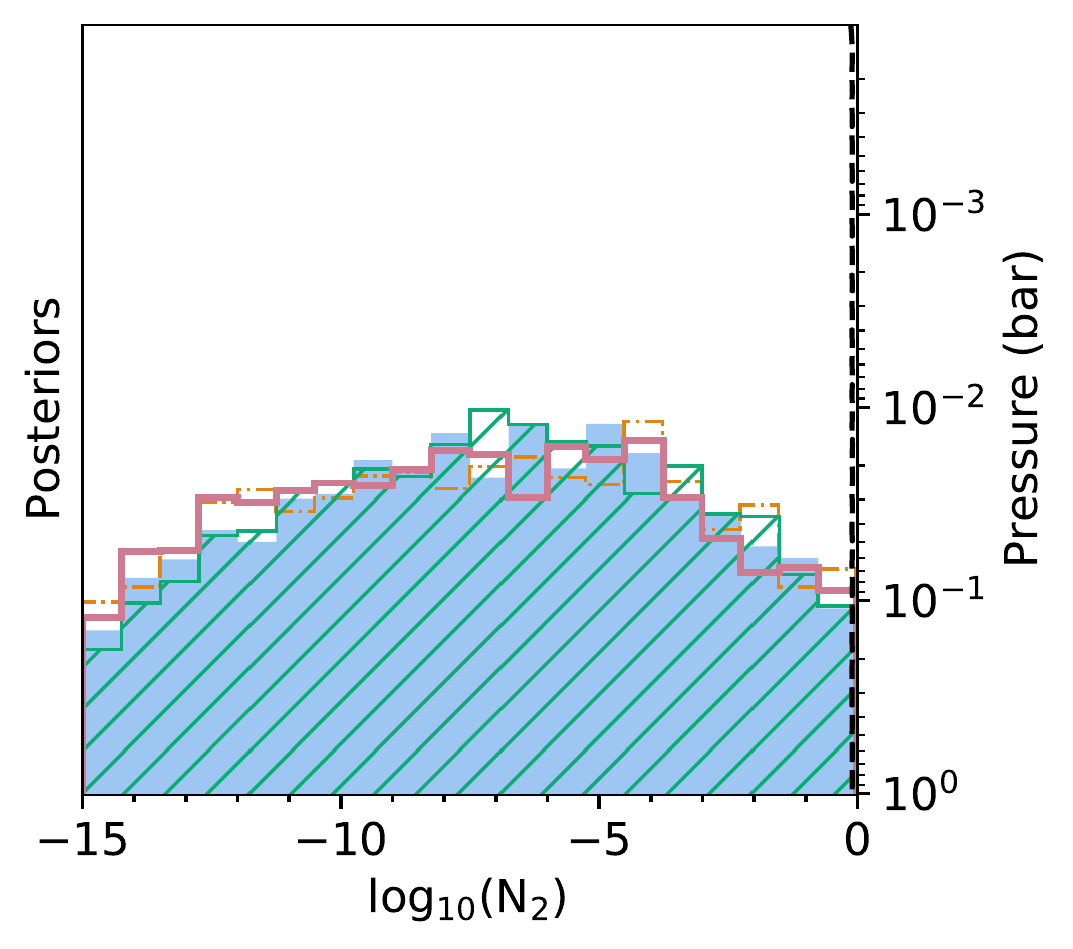}}
\,
\subfloat{\includegraphics[height=4.5cm]{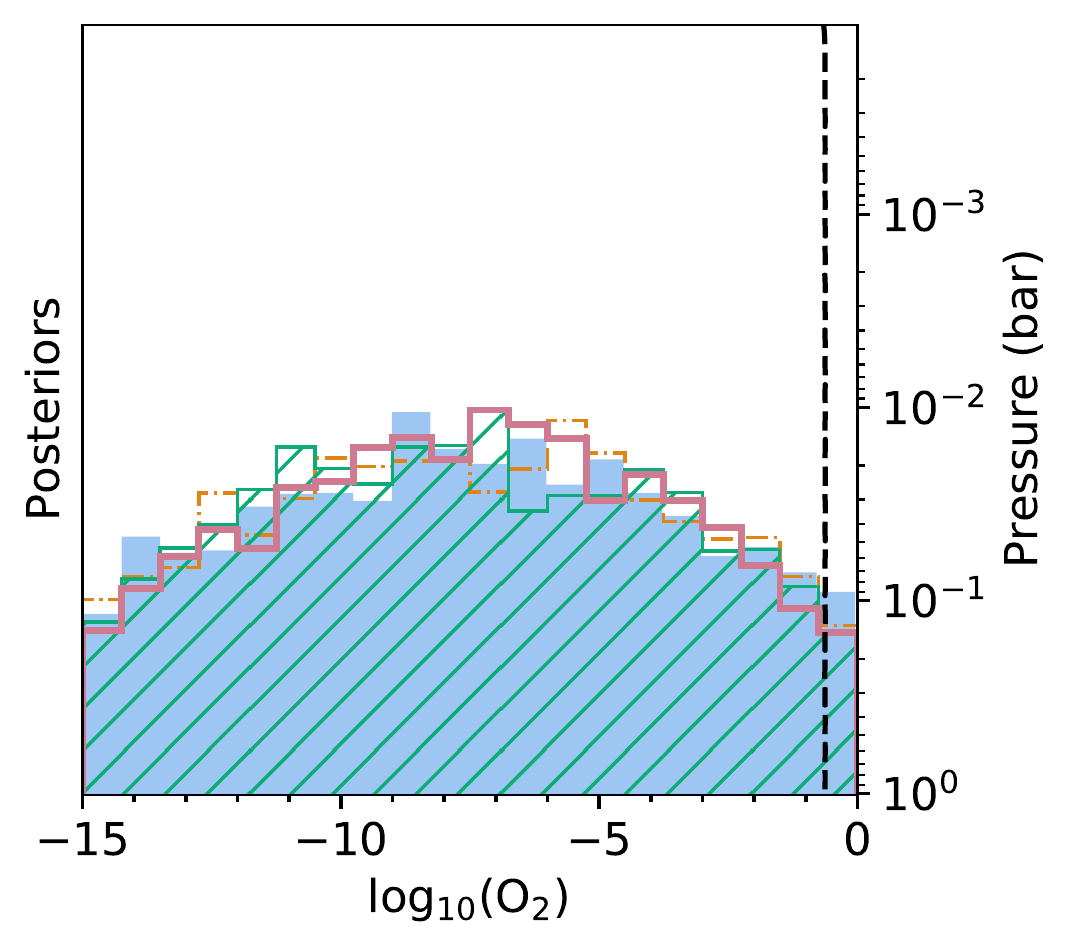}}
\,
\subfloat{\includegraphics[height=4.5cm]{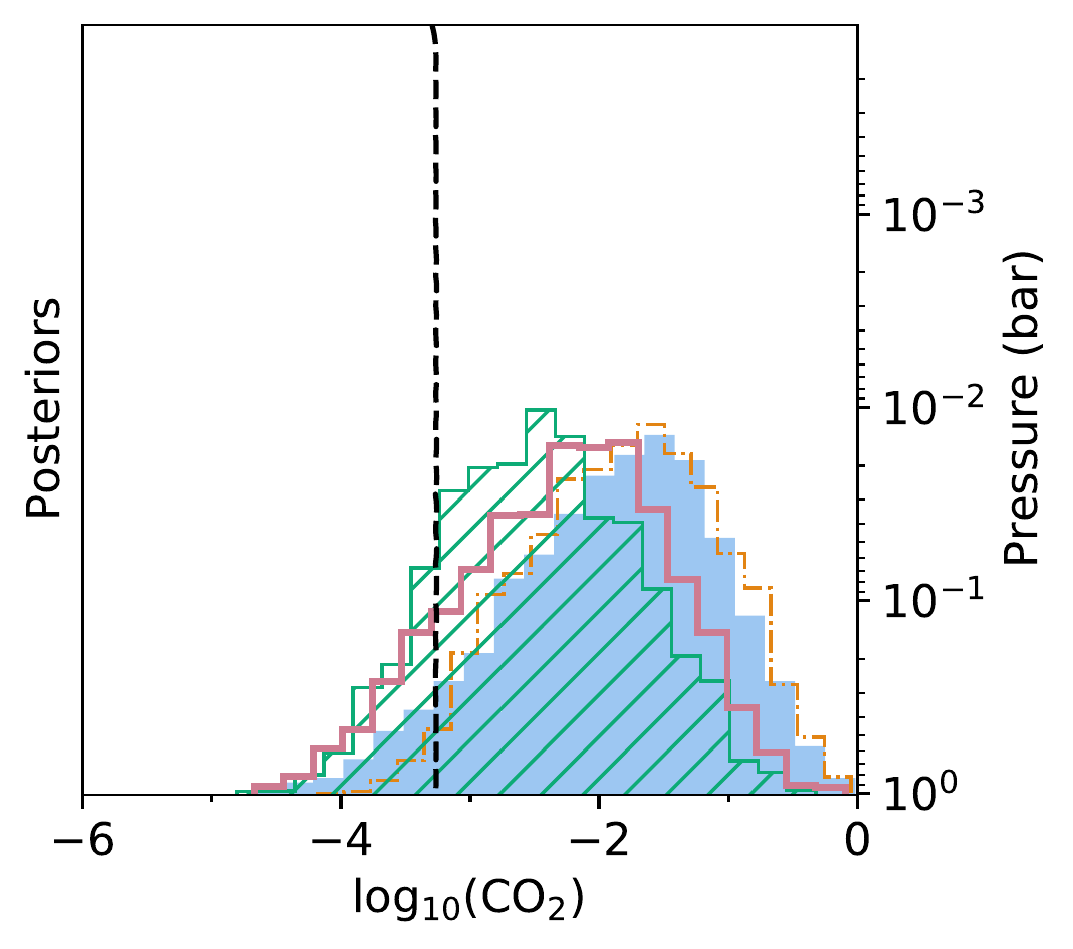}}
\\
\subfloat{\includegraphics[height=4.5cm]{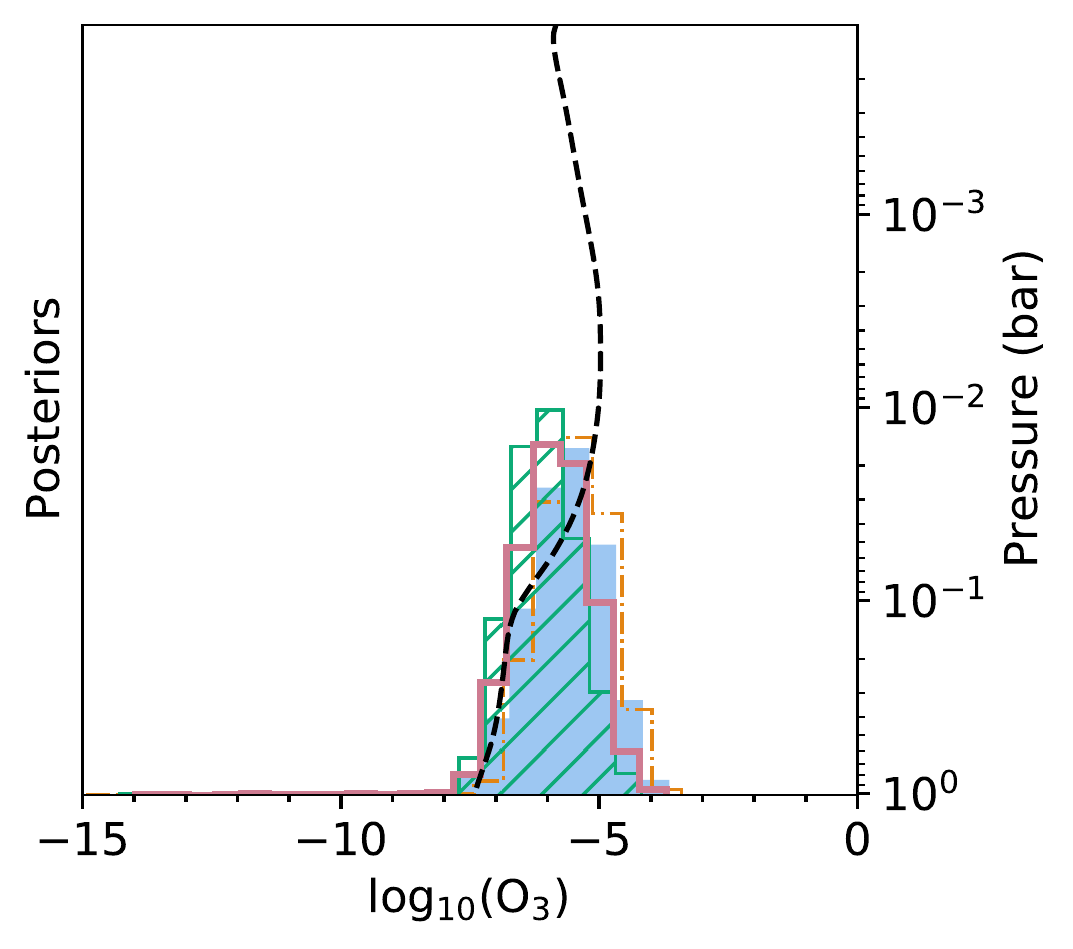}}
\,
\subfloat{\includegraphics[height=4.5cm]{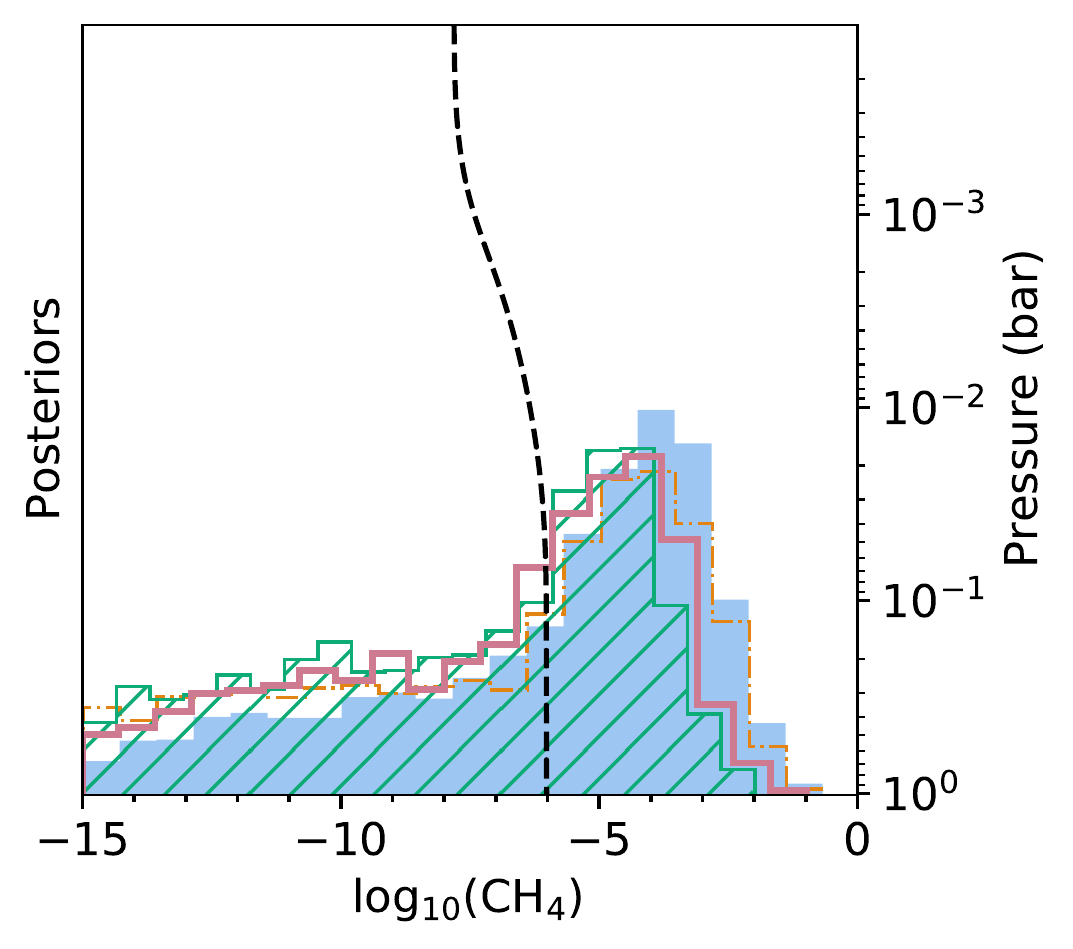}}
\,
\subfloat{\includegraphics[height=4.5cm]{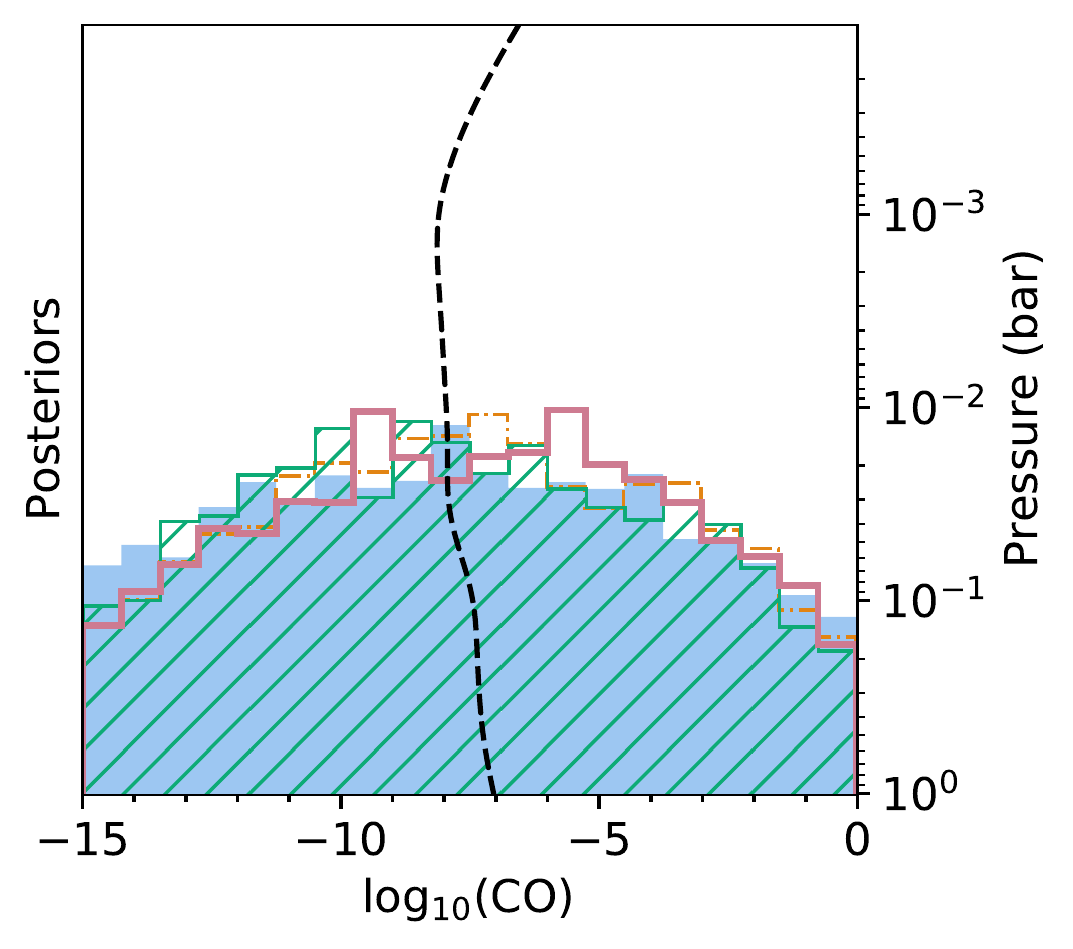}}
\\
\subfloat{\includegraphics[height=4.5cm]{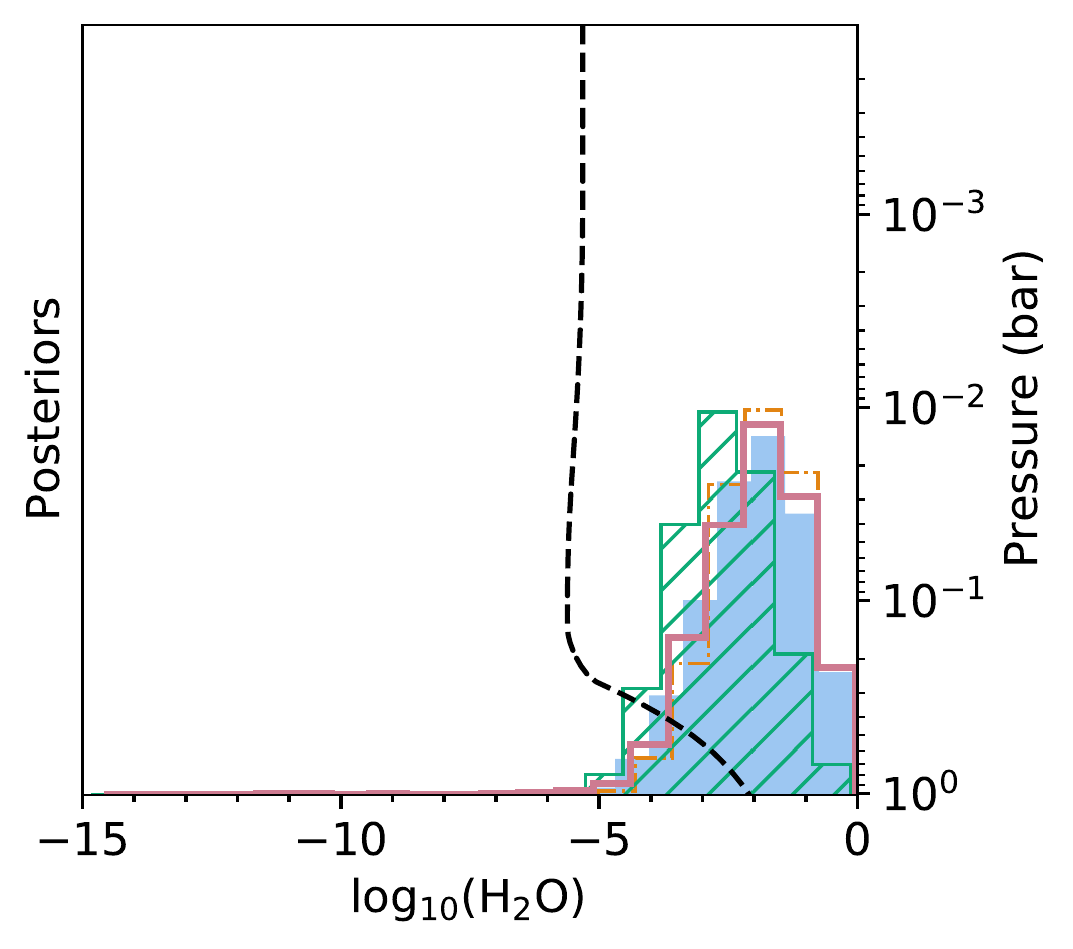}}
\,
\subfloat{\includegraphics[height=4.5cm]{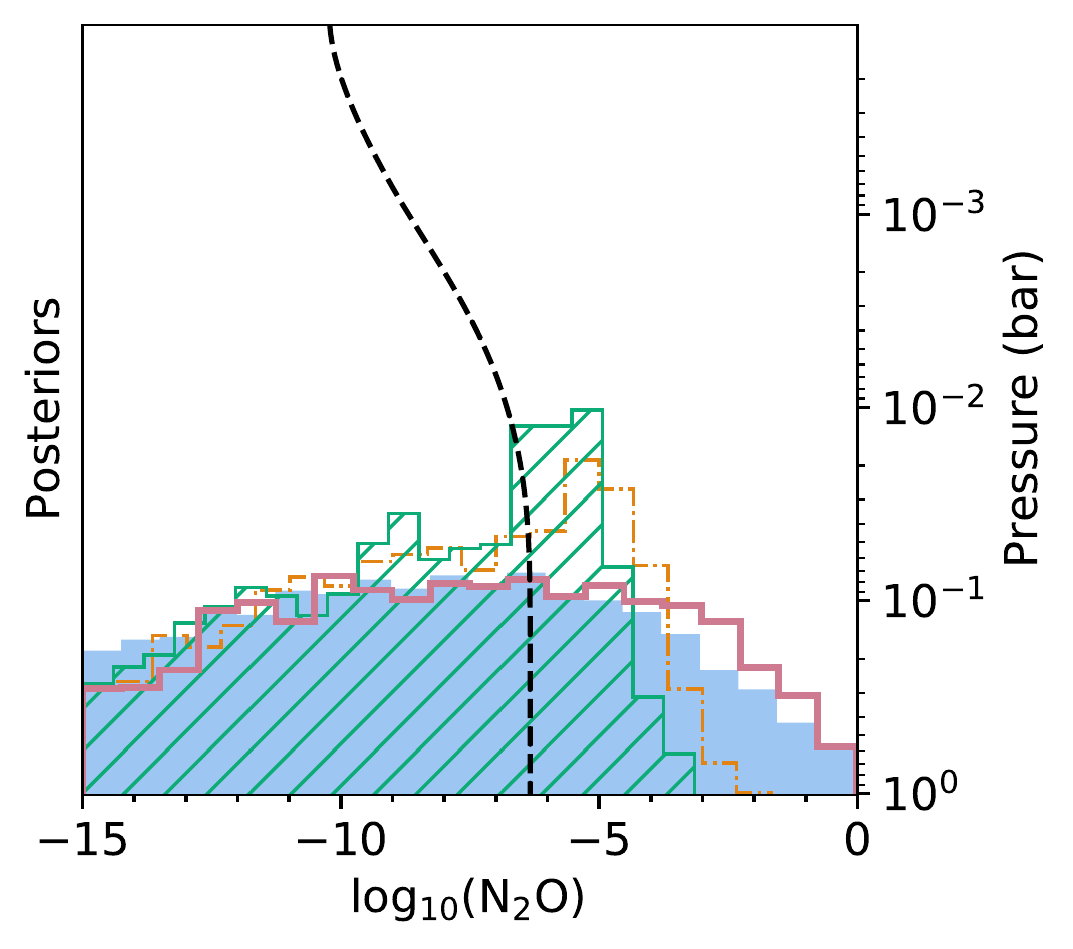}}
\,
\subfloat{\includegraphics[height=5cm]{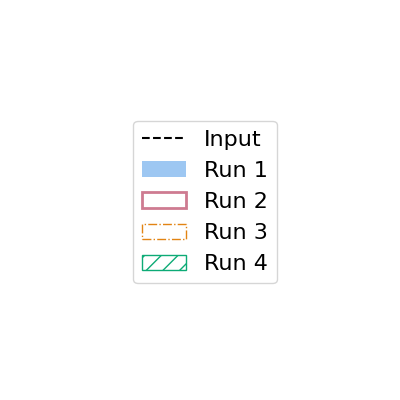}}
\\

 \caption[Posteriors] 
%>>>> use \label inside caption to get Fig. number with \ref{}
   { \label{fig:posteriors_atm} 
Posterior density distributions for the retrieved atmospheric composition for the four runs. As dashed black lines, we represent the pressure-dependent input profiles computed by Rugheimer \& Kaltenegger (2018)\cite{Rugheimer2018}. The secondary axis on the right of each subplot shows the pressure grid for the input profiles.}

\end{figure}

We now compare the performances of the four runs by analyzing the retrieved posterior density distribution for the rest of the parameters (the bulk parameters in Figure~\ref{fig:posteriors_bulk} and the atmospheric composition in Figure~\ref{fig:posteriors_atm}).

For what concerns the planetary mass and radius, there is no noticeable variation among the various runs. All runs constrain the radius with a smaller uncertainty compared to the prior. On the other hand, there is no additional constraint on the estimate of the planetary mass for any of the runs: the posterior is consistent with the Gaussian prior that was assumed for the mass. These results are consistent with our previous studies\cite{konrad2021large,Alei2022}.

The ground pressure subplot of Figure~\ref{fig:posteriors_bulk} allows us to compare the posteriors for $P_0$ for the various runs in a more quantitative way compared to the previous section. The posterior distribution of Run~4 (for which we have used opacity tables computed from HITRAN~2020, assuming air broadening and a wing cut-off of 25~cm$^{-1}$) gives a better estimate of the ground pressure compared to the other runs. The estimated $\mathrm{log_{10}(P_0)}$ in this case is $-0.4^{+0.4}_{-0.3}$ (also see the corner plot, Figure~\ref{fig:corner_4} in Appendix \ref{appendix}), which includes the true value within the 1-$\sigma$ interval. Run~3 (for which we have used tables computed from HITRAN~2020, assuming air broadening and a wing cut-off of 100~cm$^{-1}$) is the one that performs the worst: the true value is within 2-$\sigma$ to 3-$\sigma$ from the mean of the retrieved posteriors. This result shows that discrepancies in the choice of the wing cut-off have the largest impact in the determination of the correct surface pressure, even more than the choice of heterogeneous line list databases and/or broadening coefficients. This will be discussed further in Section~\ref{sec:discussion}.

In Figure~\ref{fig:posteriors_atm}, we show the posteriors for all the atmospheric species considered in our retrieval. In each subplot, we overlap a plot of the pressure-dependent input profile of the considered species, as computed by Rugheimer \& Kaltenegger (2018)\cite{Rugheimer2018}. These input profiles were used by the authors to calculate the input spectrum that we assume in this study.
In our retrieval framework, the abundance profile of each species is assumed constant with altitude. For this reason, we expect to retrieve an average estimate of the abundance of each species. We could also expect that this estimate would be more representative of the deeper layers of the atmosphere, which contribute more to the shape of the emission spectrum. %As visible from the retrieval of \ce{H2O}, and in some cases also of \ce{N2O} and \ce{CH4}, the results of the retrievals appear to match our intuition. 

All runs successfully retrieve \ce{CO2}, \ce{H2O}, and \ce{O3}, the strongest absorbers in the atmosphere of the modern Earth. In general, the retrieved estimates show very large variances, spanning multiple orders of magnitude. This is due to the low resolution and signal-to-noise ratio assumed for the baseline scenario (R=50, S/N=10). Increasing these would shrink the standard deviation of the posterior, as described in LIFE Paper V\cite{Alei2022}. For \ce{CO2}, Runs~1 and 3 overestimate significantly the abundance by more than one order of magnitude ($\geq 2 \sigma$ from the true value). On the other hand, the true abundance of \ce{CO2} lies within the 1-$\sigma$ interval of the posterior for the retrieved \ce{CO2} abundance performed by Run~4. This is a sign of the degeneracy between pressure and \ce{CO2} abundance, already encountered in our previous studies\cite{konrad2021large,Alei2022}: the same spectral feature can be caused by either higher abundances of the major absorbers at a lower pressure, or by a lower abundance of these absorbers at a higher pressure. Run~4, which provides the closest estimate of $P_0$ to the true value, is also the one for which the \ce{CO2} abundance is less overestimated. A similar effect can be noticed in \ce{H2O} and \ce{O3}. An evident correlation between these species and the ground pressure can be observed in the corner plots for each run (see Appendix \ref{appendix}).

The retrievals fail at detecting \ce{N2}, \ce{O2}, and \ce{CO} in all the scenarios. This is not surprising, since these molecules have very faint lines in the wavelength range of interest for LIFE, which are well embedded in the noise. 
For what concerns \ce{CH4}, we retrieve an upper limit for all the modeled scenarios. According to our posterior classification (see LIFE Paper III\cite{konrad2021large} for more details), for all retrieval runs we retrieve a \emph{sensitivity limit} on \ce{CH4}: there is a visible peak in the distribution, but it is impossible to rule out lower abundances of this molecule. This is consistent with our previous studies\cite{konrad2021large,Alei2022}.
On the other hand, \ce{N2O} was not detected in LIFE Paper V (Run~1) nor Run~2, but Runs~3 and 4 allow to rule out abundances smaller than 1-0.1\%. The posterior in Run~4 is also a sensitivity limit, according to our posterior classification scheme. Ultimately, it would be necessary to run retrievals with and without \ce{N2O} to properly assess the significance of this tentative detection. Still, this qualitative shrinking of the posterior distribution can demonstrate how the choice of the database (HITRAN~2020 instead of ExoMol, see Table~\ref{tab:setups}) could play a role in the detection of biosignature gases. We will discuss this in more detail in Section~\ref{sec:discussion}.

\section{DISCUSSION}\label{sec:discussion}

We can identify some potential sources of systematics in the absorption cross-sections used in the various runs:

\begin{itemize}
    \item \emph{Database versioning.} Line transition databases are maintained and updated periodically, as the laboratory and modeling efforts are carried out. For this reason, it is common to find discrepancies among various versions of the same database (see e.g., the latest publication by the HITRAN team\cite{GORDON2022107949}), both in terms of line transitions and spectroscopic parameters. 
    \item \emph{Various databases.} Although different databases (e.g., HITRAN and ExoMol) are generally cross-validated with one another, and with laboratory experiments or observations, there could be differences in the number of transitions and the quality of the spectroscopic data in the parameter space where some databases are not the ideal choice to use. Different databases might be more suited than others for specific case scenarios. For example, the HITRAN database is composed mainly of experimental data measured at room temperature. It is therefore considered the most accurate database in the temperate range (such as an Earth-twin, which we model in this study). On the other hand, theoretical calculations are needed to correctly calculate transition lines at higher energies: ExoMol\cite{Chubb2021,Tennyson2020} would then be more appropriate to use in this case.
    \item \emph{Broadening coefficients.} The lack of knowledge about some spectral parameters (such as broadening coefficients) leads to further differences. In this case, the modeling of the line broadening must rely on approximations and assumptions. An example of this is the \ce{CH4} opacity table provided by \texttt{petitRADTRANS}\cite{Molliere2019} (used in Run~1), which uses Eq. 15 from Sharp \& Burrows (2007)\cite{Sharp2007}, an analytical formula based on the rotational quantum number. Albeit driven by common sense, some of these assumptions are arbitrary and are used differently in the astronomical community. Broadening coefficients for H-He and ``air'' (based on the composition of modern Earth's atmosphere) are commonly assumed in the community, but they might not be adequate in a few case scenarios (e.g., \ce{H2O}-dominated atmospheres on high-metallicity exoplanets\cite{Ehsan2019}, where \ce{H2O} self-broadening would need to be considered).
    \item \emph{Wing cut-off.} A common technique used in the community is the hard cut-off at 25~cm$^{-1}$ from the line core. However, many radiative transfer models use different cut-off functions. For example, the default \texttt{petitRADTRANS} opacities\cite{Molliere2019} (see Run~1 in Table~\ref{tab:setups}) assume a sub-Lorentzian cut-off instead of a step-function at a given distance from the line cut-off. On the other hand, opacities produced by ExoMolOP\cite{Chubb2021} for \texttt{petitRADTRANS} (used in Run~2), use Eq. 6 from Chubb et al. (2021)\cite{Chubb2021}, which is a hard cut-off but whose threshold varies depending on pressure, temperature, and wavelength.  As shown in Baudino et al. (2017)\cite{2017ApJ...850..150B}, discrepancies in the line cut-off can cause noticeable differences in the modeled spectra, especially in the emission peak. 
    \item \emph{Isotope transitions}. Line lists databases also include isotopic transitions of a given species. Including these transitions in the computation of line opacity tables might therefore play a role, especially when isotopologues are relatively abundant in the atmosphere. In Earth's case, the \ce{^13C}-based isotopes of \ce{CO2}, \ce{CH4}, and \ce{CO} compose about 1\% of the total abundance of each species. This aspect was not taken into account in this study and only reported here for completeness. 

\end{itemize}

    \begin{figure} [ht]
   \begin{center}
   \begin{tabular}{c} %% tabular useful for creating an array of images 
  \subfloat{\includegraphics[width=0.45\textwidth]{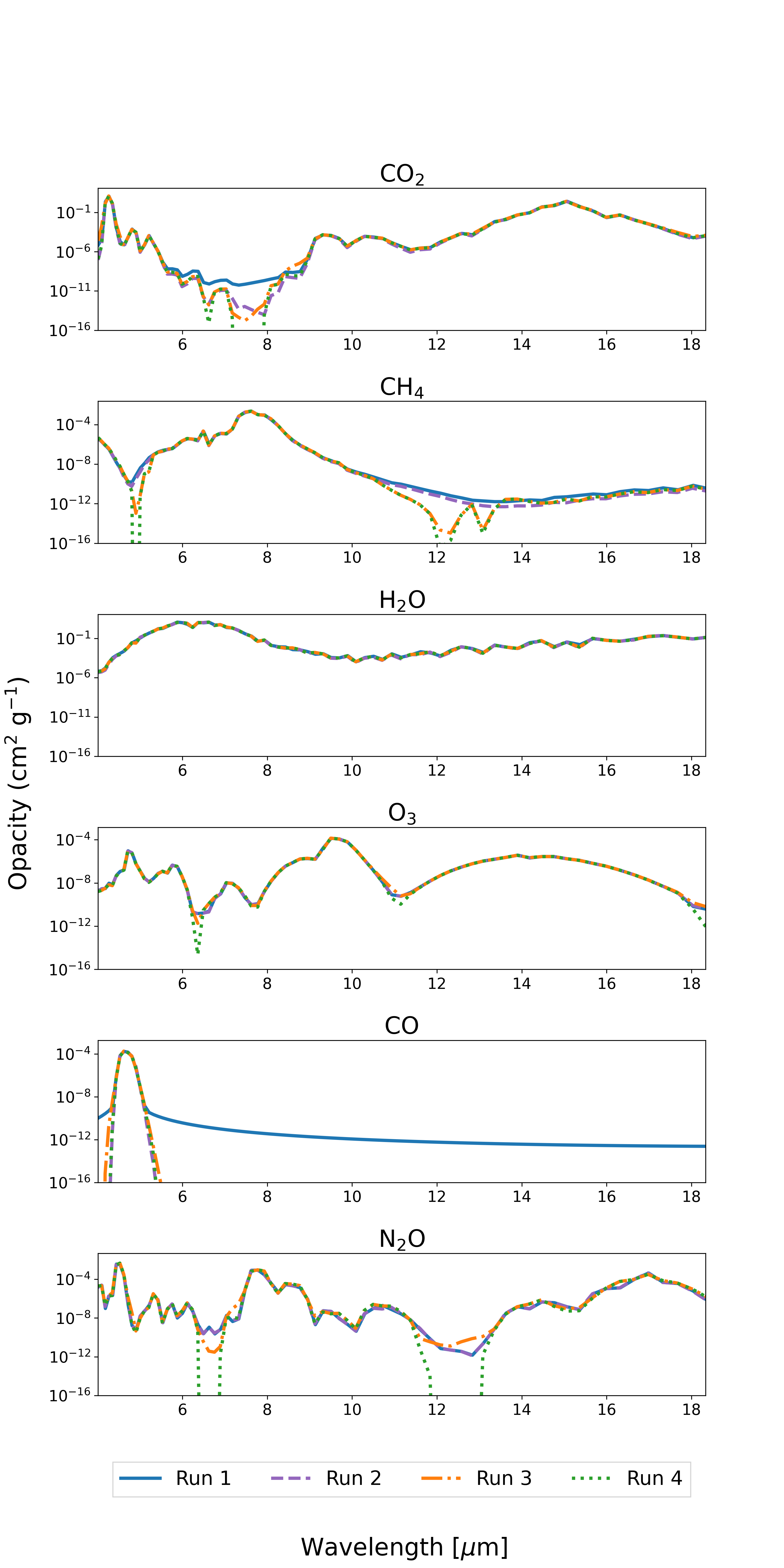}}
\,
\subfloat{\includegraphics[width=0.45\textwidth]{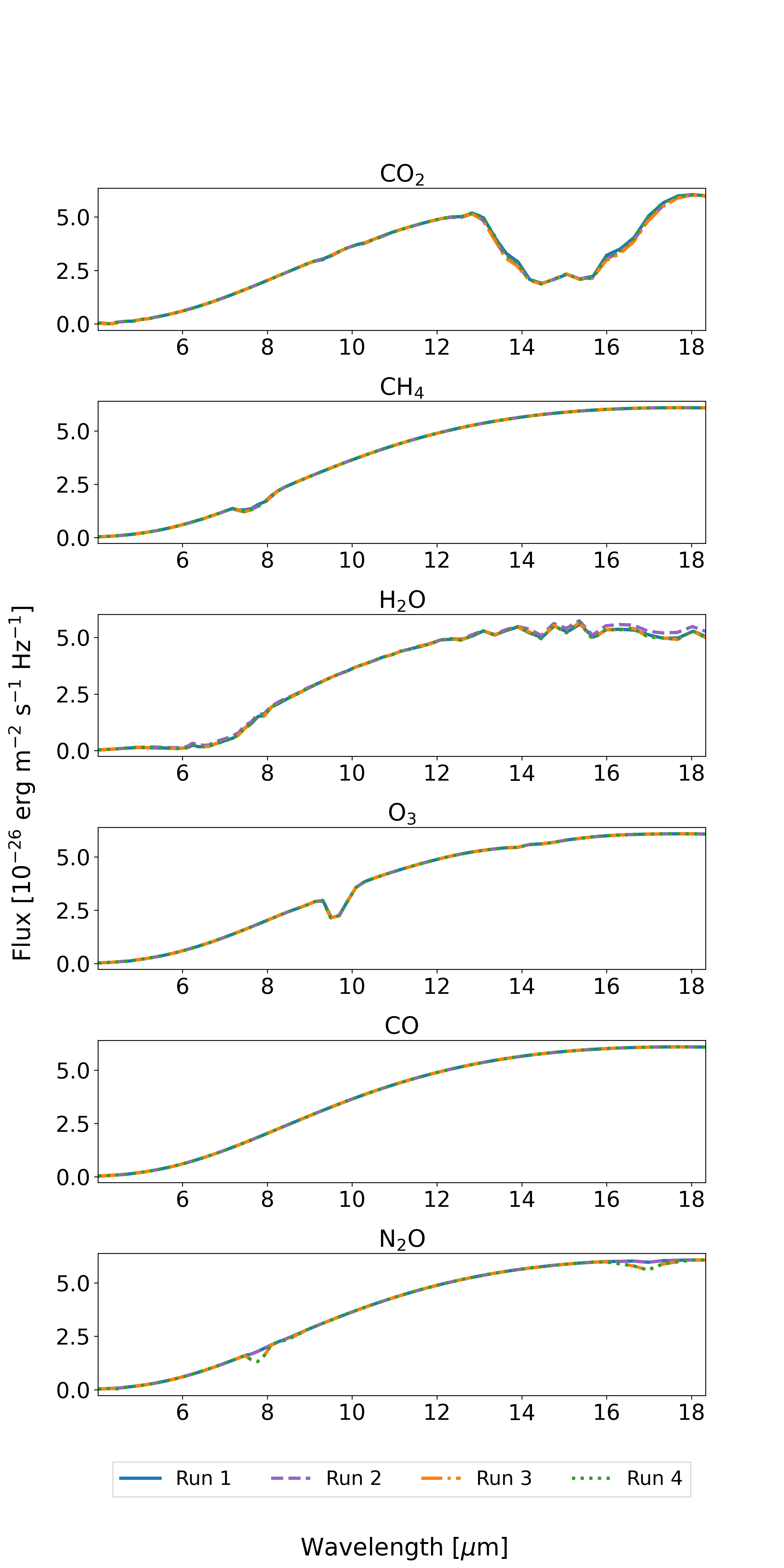}}
\\
   \end{tabular}
   \end{center}
   \caption[] 
%>>>> use \label inside caption to get Fig. number with \ref{}
   { \label{fig:lines2} 
Comparison of the various opacities and their impact on the emission flux. \emph{Left column}: Line opacity of the deepest atmospheric layer (288 K and 1 bar) at a resolution R=50, for all the absorbers in the atmosphere (rows), assuming the atmospheric composition at the deepest layer for the four runs (color-coded according to Table~\ref{tab:setups}). \emph{Right column}: emission spectrum of the Earth's atmosphere considering the absorption of only one species (different species in different rows), at R=50, assuming the input profiles for the temperature and the abundances computed in Rugheimer \& Kaltenegger (2018)\cite{Rugheimer2018} for the four runs (color-coded according to Table~\ref{tab:setups}).}
   \end{figure} 

In the left column of Figure~\ref{fig:lines2}, we compare the opacities of the deepest layer of the atmosphere (at p = 1 bar, T = 288 K) for each absorber (rows) and the four retrieval runs (color-coded based on Table~\ref{tab:setups}). The opacities are weighted for the abundance of each species in the deepest layer of the atmosphere.
In general, the line cores (high opacity regions) are in strong agreement among the various tables, meaning that the line transition databases (HITEMP, ExoMol, HITRAN~2020) are consistent with one another. 
Larger differences can be seen in the regions at low opacity, where the line wings dominate. Different choices in the cut-offs translate into a different contribution of the line wings farther away from the line cores. The default opacity tables assumed in \texttt{petitRADTRANS} (the ones considered for Run~1), are generally more absorptive in the line wings (e.g. in the case of \ce{CO2} between 6-8 $\mu$m, \ce{CH4} between 11-14 $\mu$m, and \ce{CO} in most of the wavelength range). This is a result of a sub-Lorentzian cut-off, characterized by an exponential decrease, compared to the hard cut-offs used in the other opacity tables.

In the right column of Figure~\ref{fig:lines2}, we show the emission flux of an Earth-twin assuming only one absorber in the atmosphere. For each species (rows), we consider the input abundance profile and the input P-T profile calculated by Rugheimer \& Kaltenegger (2018)\cite{Rugheimer2018}. Here, the differences between the various emission spectra are less noticeable. Some differences appear on the emission spectrum that assumes \ce{H2O} as an absorber: here, the ExoMol \ce{H2O} (Run~2) opacity tables cause the total emission to be slightly higher ($<10\%$ increase compared to the other runs) at wavelengths higher than 14 $\mu$m. For what  concerns \ce{N2O}, both the opacities used in Runs~1 and 2 do not show any significant spectral feature, while the opacity tables based on HITRAN~2020 (Runs~3 and 4) show the absorption of the \ce{N2O} lines at about 8 and 17 $\mu$m. It was probably the presence of more spectral lines that allowed a qualitative detection of \ce{N2O} in Runs~3 and 4 (with an upper limit and a sensitivity limit respectively), compared to an unconstrained posterior for this species in Runs~1 and 2. 
\ce{CH4} and \ce{CO} don't cause detectable variations in the spectrum, even though the opacity tables for these two species appeared different among the various runs (see left column of Figure~\ref{fig:lines2}). This is justified since these species are less abundant in the atmosphere and absorb at shorter wavelengths ($\lambda<6\ \mu$m), for which the thermal emission is lower. 

Such discrepancies in the opacity tables are the reason for the slight difference in the retrieval results that we described in Section~\ref{sec:results}.
While we cannot fully disentangle the contribution of the four sources of systematics that we identified earlier in this section in the four retrieval runs, we can statistically quantify which model might perform better in retrieving the simulated observation. 

To do this, we can calculate the Bayes' factor between each pair of models. This is a technique that is widely used in the community  to identify the setup that best fits the data. It is defined by the following equation:

\begin{equation}\label{eq:bayes}
 K=\mathcal{Z}_{\mathrm{M}_1}/\mathcal{Z}_{\mathrm{M}_2}
\end{equation}

\begin{figure} [ht]
   \begin{center}
   \begin{tabular}{c} %% tabular useful for creating an array of images 
   \includegraphics[width=\textwidth]{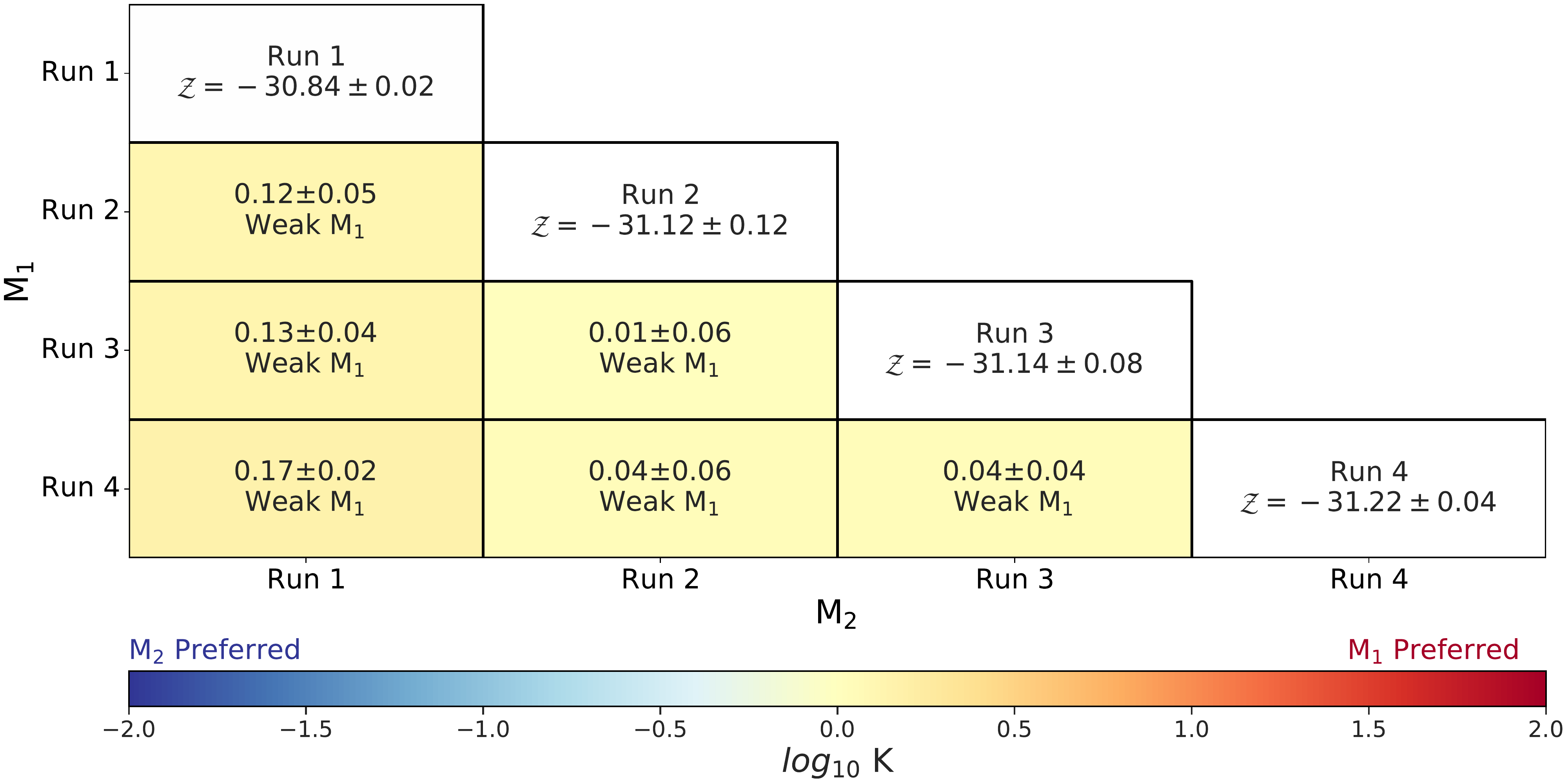}
   \end{tabular}
   \end{center}
   \caption[example] 
%>>>> use \label inside caption to get Fig. number with \ref{}
   { \label{fig:bayesianevidence} Corner plot showing the Bayes' factors and their evidence strength according to the Jeffrey's scale (Table~\ref{table:jeffrey}) for every pair of runs, color-coded according to the color bar on the bottom. In the diagonal cells, we showed the values of the Bayesian evidence for each run.
}
   \end{figure}

Where ${Z}_{\mathrm{M}_i}$ are the Bayesian evidences of a model $\mathrm{M}_i$, for every pair of retrieval setups $\mathrm{M}_1$ and $\mathrm{M}_2$:

\begin{table}[h]
\caption{Jeffrey's scale \cite{Jeffreys:Theory_of_prob} for the interpretation of the Bayes' factor $K=\mathcal{Z}_{\mathrm{M}_1}/\mathcal{Z}_{\mathrm{M}_2}$. Adapted from LIFE Paper III\cite{konrad2021large}.}
\label{table:jeffrey} 
\centering 
\begin{tabular}{|l|l|}
\hline
$\mathbf{\log_{10}\left(K\right)}$ & \textbf{Evidence Strength}\\ 
\hline 
$(-\infty,-2]$ &Decisive support for $\mathrm{M}_2$\\\hline
$(-2,-1]$ &Strong support for $\mathrm{M}_2$\\\hline
$(-1,-0.5]$ &Substantial support for $\mathrm{M}_2$\\\hline
$(-0.5,0]$ &Very weak support for $\mathrm{M}_2$\\\hline
 $(0,0.5)$ &Very weak support for $\mathrm{M}_1$\\\hline
 $[0.5,1)$ &Substantial support for $\mathrm{M}_1$\\\hline
 $[1,2)$ &Strong support for $\mathrm{M}_1$\\\hline
 $[2,\infty)$ &Decisive support for $\mathrm{M}_1$\\\hline

\end{tabular}
\end{table}
Different values of $K$ can be interpreted in terms of evidence strength by using Jeffrey's classification (Table~\ref{table:jeffrey}).
For every combination of retrieval runs, we computed the Bayes' factor and retrieved its classification. We show the result of this analysis as a corner plot in Figure~\ref{fig:bayesianevidence}. The differences among the various runs are generally small: in all cases, Jeffrey's classification suggests only a weak preference for one model over the other, for every combination. Still, Runs~2, 3, and 4 outperform Run~1 (see the first column of the corner plot). This is an interesting result, which could imply that the results of the retrievals run in LIFE Paper V might be improved by varying the opacity tables. Runs~2, 3, and 4 are not particularly preferred with respect to one another: the Bayes' factor is very close to 0, considering its uncertainty, for all combinations.

We recall that all of the opacity tables used in the four retrieval runs were different from the ones used by Rugheimer \& Kaltenegger (2018)\cite{Rugheimer2018} to compute the input spectrum.
The most similar tables were the ones used in Run~4, the only difference being the versioning of the same line transition database (HITRAN~2020 versus HITRAN~2016). For this reason, this retrieval run retrieves better results than the others, both in terms of Bayes' factor and parameter estimation.

The opacity tables used in the other runs differ from the ones used by Rugheimer \& Kaltenegger (2018)\cite{Rugheimer2018} in terms of database provenance, versioning, broadening, and wing cut-off. 
The different treatment of the wing cut-off causes the line wings to contribute differently to the spectrum, especially in the wavelength bands where there is little to no absorption. This has an impact on the retrieval of the ground pressure, which in turn can bias the correct characterization of the atmospheric components. 

We argue that the different wing cut-off in the opacity tables is the main reason for the lower performance of Run~1 compared to the others. The opacity tables used in Run~1 were the only ones that assumed a sub-Lorentzian wing cut-off, compared to the different but nevertheless hard step-function cut-offs of the opacities used in the other runs. Meanwhile, the input spectrum was also calculated using opacity tables that assumed a hard cut-off at 25~cm$^{-1}$ from the line core. Because of the sub-Lorentzian cut-off, the performance of Run~1 could have been reduced, while the runs that assumed hard wing cut-offs, albeit at different thresholds, performed nearly equally as well.

From the results presented in Section~\ref{sec:results}, as well as the Bayes' factor analysis, we can assume that the choice of the ``appropriate" opacity table won't impact significantly the quality of the retrieval of a simulated cloud-free Earth at the current baseline requirements for LIFE (see Table~\ref{tab:lifesim}). 
This might not be the case for other scenarios. As shown in Gharib-Nezhad \& Line (2019)\cite{Ehsan2019}, different assumptions on the broadening coefficients can cause differences up to 100s ppm in the transmission spectrum of sub-Neptunes/super-Earths, which are not negligible considering the precision of JWST and other future telescopes.

Retrievals of high-resolution observations might be even more impacted by the choice of the correct opacities (see e.g., Rocchetto 2017\cite{Rocchetto2017}). Tannock et al. (2022)\cite{Tannock2022} showed that having accurate and up-to-date \ce{CH4} and \ce{H2O} line lists is necessary to correctly fit a high-resolution spectrum of a cold brown dwarf (T$\approx 1000\ K$) obtained with IGRINS. In this work, they suggest using the HITEMP line lists for \ce{CH4}, now included in the HITRAN~2020 database, and the ExoMol opacities for \ce{H2O}. This is likely an aspect that should also be kept in mind when designing future high-resolution spectrographs.

It is worth pointing out that this does not necessarily mean that some opacity tables rather than others are to be preferred in any future retrieval effort. The ``goodness" of each opacity table varies depending on the input spectrum to be analyzed.  Ideally, when retrieving real observed spectra, we should perform multiple retrievals assuming different retrieval setups and then compare the evidence to test the robustness of each run (through the analysis of the Bayes' factor or other statistically significant techniques).

The radiative transfer models are also limited by simplifications and assumptions. In order to perform analyses on real observations, a correct treatment of clouds, 3D effects, and disequilibrium temperature should also be taken into account, to reduce biases in retrievals\cite{BarstowHeng}. As noted by Rocchetto (2017)\cite{Rocchetto2017} and Barstow et al. (2022)\cite{Barstow2022}, not only the line lists and spectral parameters play a role in the atmospheric retrievals, but also the sampling of the opacities to compute correlated-k absorption cross-sections. 
To understand and quantify such limitations, inter-model comparison studies are being carried out by comparing the most used radiative transfer models with one another and as forward models of retrieval routines. Notable examples are Barstow et al. (2020; 2022)\cite{Barstow2020,Barstow2022} and the current studies performed within the Climates Using Interactive Suites of Intercomparisons Nested for Exoplanet Studies (CUISINES) NExSS Working Group\cite{Fauchez2021}. 

The parameter estimation routines used in Bayesian retrievals also have to rely on simplifications due to the limit on the number of parameters to ensure a reasonable computing time. As described in more detail in our previous studies\cite{konrad2021large,Alei2022}, our retrieval framework assumes a pressure-constant abundance profile for all the species in the atmosphere, but this is not a realistic representation of the atmosphere of modern Earth, nor any other atmosphere studied so far. 
Furthermore, in the case of future missions and concepts such as LIFE, we have to make assumptions about the noise levels and distribution. As described in LIFE Paper II\cite{Dannert2022}, our calculation only considers fundamental noise, assuming that it dominates the instrumental noise which can therefore be neglected. We additionally assume that the noise is independent and identically distributed and follows a normal distribution. However, non-Gaussian noise might also be a source of systematics in retrievals, as described in Ih \& Kempton (2021)\cite{Ih2021}.

All in all, atmospheres are very complex and we will be constantly limited by the unknowns. Still, it is important to be aware of all the possible sources of systematics that play a role in atmospheric retrievals and to minimize them as much as possible. On the other hand, laboratory and modeling efforts should continue, to constantly update our knowledge about line transitions and spectral parameters and to better prepare for upcoming observations\cite{Fortney2019}.

\section{Summary and future work} \label{sec:conclusion}

In this study, we performed a deeper analysis on the impact of the choice of absorption cross-sections in the retrieval of an Earth-twin orbiting a Sun-like star at a distance of 10 pc from the observer. We assumed to detect such an exoplanet with LIFE (the Large Interferometer For Exoplanets). Still, such a study should also be useful for both current (like JWST) and future missions (like the LUVOIR-HabEx concept). 

We performed four different retrievals on the same input spectrum (computed by Rugheimer \& Kaltenegger 2018\cite{Rugheimer2018}), changing the opacity tables that the forward model was allowed to use (see Table~\ref{tab:setups} for more details). We varied the provenance of the line lists (various databases and different versions of the same database), the broadening coefficients, and the line wing cut-off functions. We found that using a heterogeneous set of opacity tables in terms of provenance, broadening, and cut-off (Run~1) has the lowest performance compared to the other runs. On the other hand, the retrieval run that used the most similar opacities to the ones used to produce the input spectrum (Run~4) slightly outperforms the other runs, particularly in estimating the ground pressure and the abundances of the major absorbers in the atmosphere. Runs~2, 3, and 4 also allow us to retrieve at least an upper limit on \ce{N2O}, which is a relevant biosignature gas. 
We argue that the largest impact on the retrievals is caused by a different treatment of the wing cut-off in the opacity tables. The wing cut-off affects the opacity of the spectral continuum, which is the region that is most relevant for retrieving the ground pressure $P_0$. Because of the pressure-abundance degeneracy, the incorrect retrieval of $P_0$ also biases the retrieval of the chemical composition (especially \ce{CO2}, one of the major atmospheric absorbers).

From this study, it appears clear that correct treatment of the absorption cross-sections is necessary to perform retrievals. This translates into the need for continued laboratory and modeling studies to help fill up the knowledge gaps in the spectral parameters. To be able to correctly reproduce a wider variety of scenarios, we are computing updated opacity tables that could be used for \texttt{petitRADTRANS} assuming a larger set of broadening parameters and wing cut-off functions. These will soon be included in the default set of correlated-k tables for this model.

Furthermore, radiative transfer models and Bayesian retrieval frameworks used in the community should be compared with one another and further characterized, to identify similarities and differences. This effort is currently carried out in the BUFFET program within the CUISINES NExSS Working Group\cite{Fauchez2021}. These inter-model comparisons will also include \texttt{petitRADTRANS}\cite{Molliere2019} and the Bayesian retrieval routine used by our group\cite{konrad2021large,Alei2022}. 

Real data will be more complex than any theoretical model, which is limited by unknowns and necessary approximations. To build the next generation of instruments and to prepare for future observations, it is of paramount importance to identify and minimize the systematic errors, as well as to identify critical points for the community to work on in the next decades.

\acknowledgments % equivalent to \section*{ACKNOWLEDGMENTS}       
This work has been carried out within the framework of the National Centre of Competence in Research PlanetS supported by the Swiss National Science Foundation under grants 51NF40\_182901 and 51NF40\_205606. S.P.Q. and E.A. acknowledge the financial support of the SNSF. P.M. acknowledges support from the European Research Council under the European Union’s Horizon 2020 research and innovation program under grant agreement No.~832428.
% References
\bibliography{biblio} % bibliography data in report.bib
\bibliographystyle{spiebib} % makes bibtex use spiebib.bst

\appendix
\section{Ancillary Plots}\label{appendix}

 \begin{figure} [ht]
   \begin{center}
   \begin{tabular}{c} %% tabular useful for creating an array of images 
   \includegraphics[width=\textwidth]{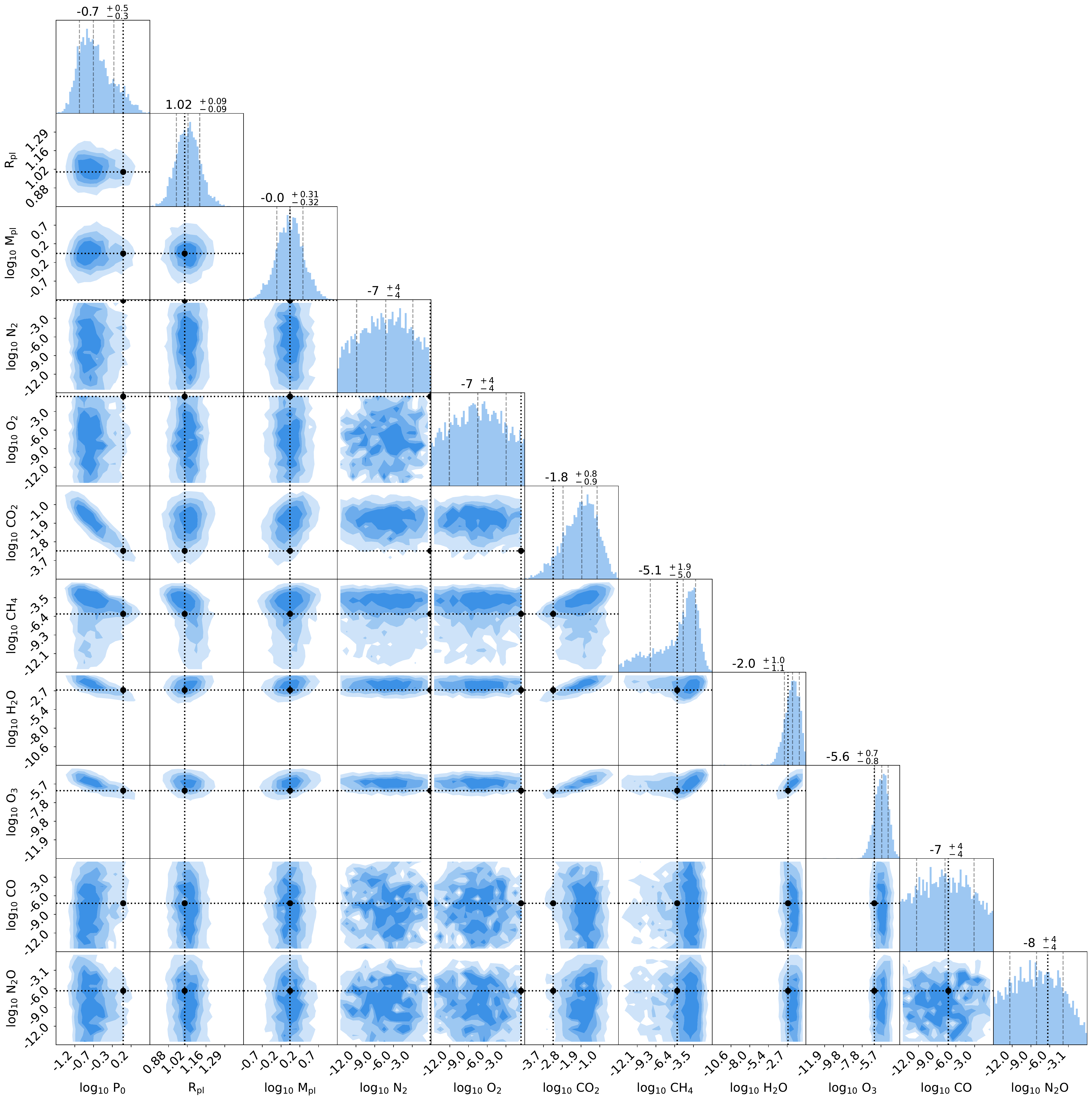}
   \end{tabular}
   \end{center}
   \caption[example] 
%>>>> use \label inside caption to get Fig. number with \ref{}
   { \label{fig:corner_1} Corner plot for the posterior distributions from the retrievals of Run~1 (color-coded according to Table~\ref{tab:setups}). The black lines indicate the expected values for every parameter. On the diagonal, we show the marginalized posteriors for each parameter, with median and 1-$\sigma$ uncertainties as dashed gray lines. 
}
   \end{figure}

    \begin{figure} [ht]
   \begin{center}
   \begin{tabular}{c} %% tabular useful for creating an array of images 
   \includegraphics[width=\textwidth]{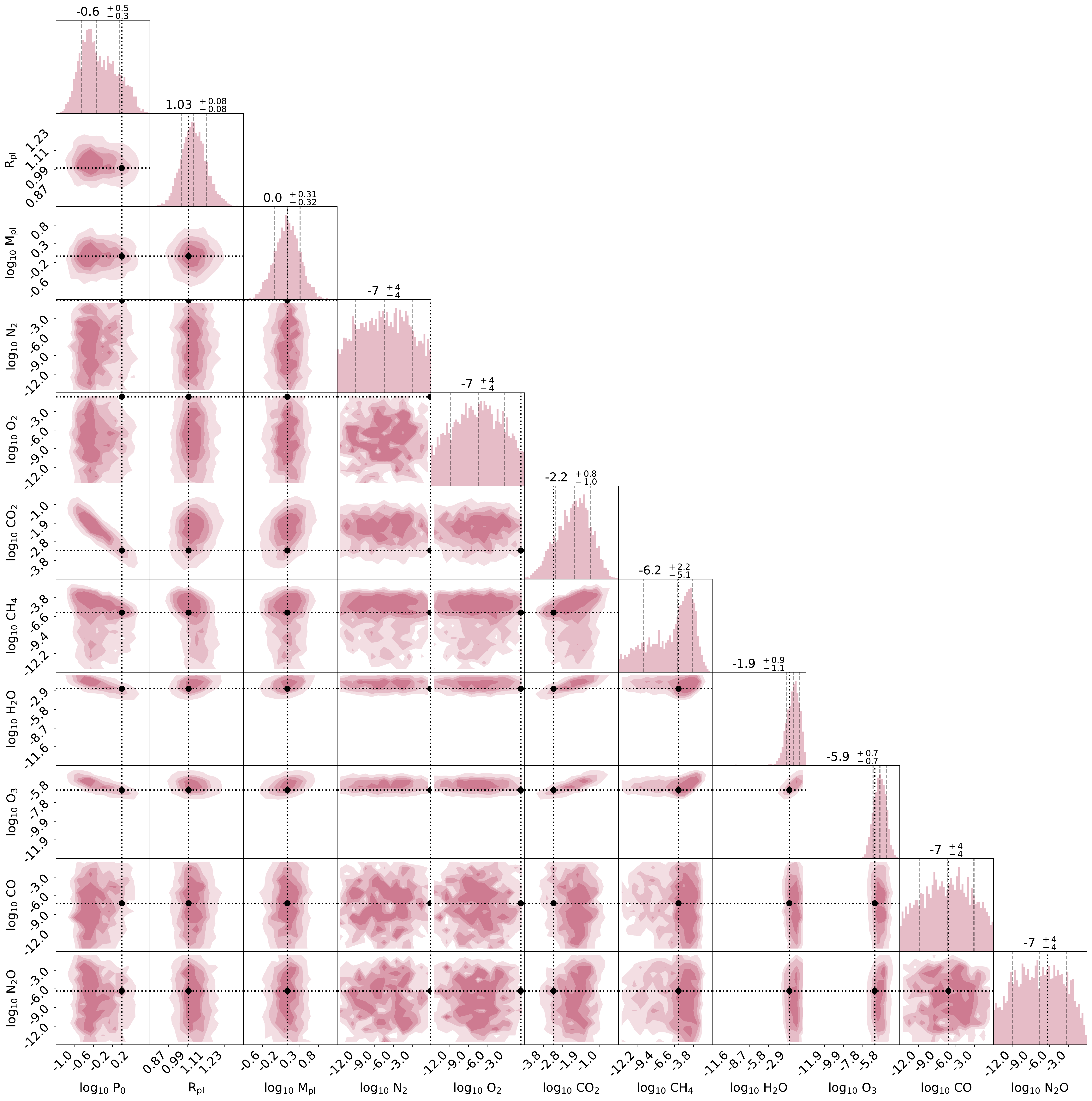}
   \end{tabular}
   \end{center}
   \caption[example] 
%>>>> use \label inside caption to get Fig. number with \ref{}
   { \label{fig:corner_2} Same as Figure~\ref{fig:corner_1}, but for Run~2.
}
   \end{figure} 
   
    \begin{figure} [ht]
   \begin{center}
   \begin{tabular}{c} %% tabular useful for creating an array of images 
   \includegraphics[width=\textwidth]{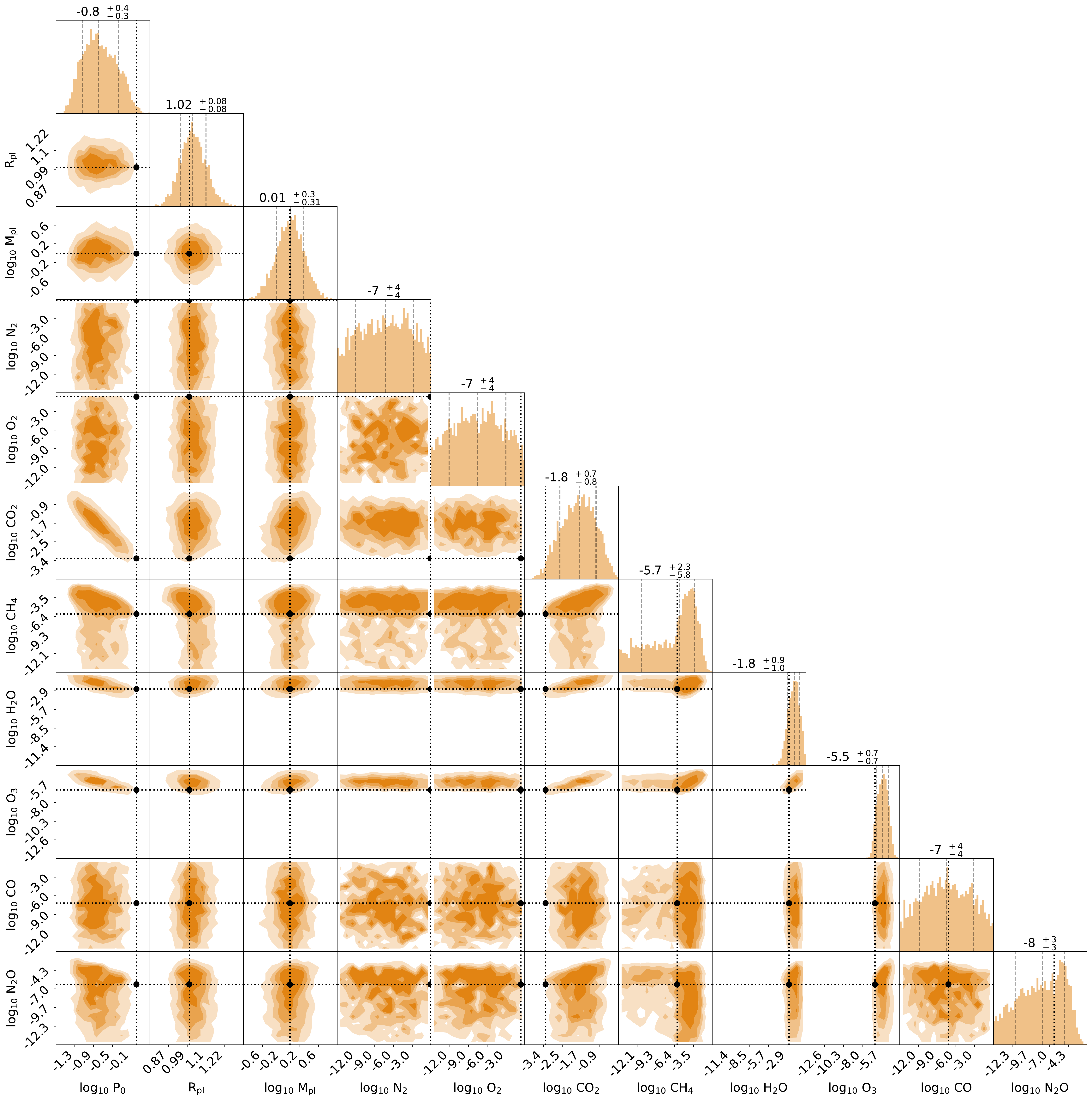}
   \end{tabular}
   \end{center}
   \caption[example] 
%>>>> use \label inside caption to get Fig. number with \ref{}
   { \label{fig:corner_3} Same as Figure~\ref{fig:corner_1}, but for Run~3.
}
   \end{figure} 
   
    \begin{figure} [ht]
   \begin{center}
   \begin{tabular}{c} %% tabular useful for creating an array of images 
   \includegraphics[width=\textwidth]{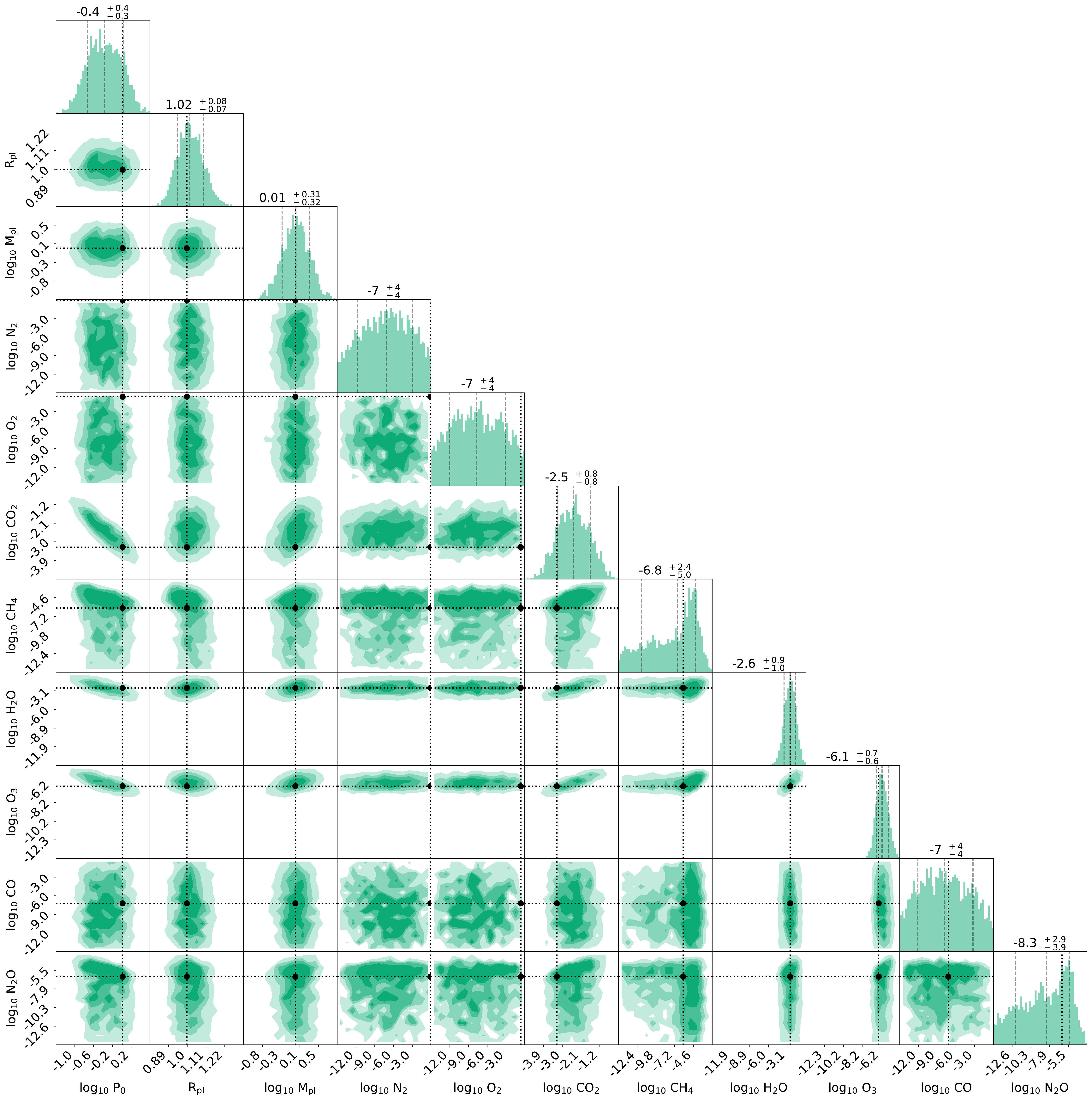}
   \end{tabular}
   \end{center}
   \caption[example] 
%>>>> use \label inside caption to get Fig. number with \ref{}
   { \label{fig:corner_4} Same as Figure~\ref{fig:corner_1}, but for Run~4.
}
   \end{figure} 
   
\end{document}